Title

# Timing of the faulting on the Wispy Terrain of Dione based on stratigraphic relationships with impact craters

Authors

**Naoyuki Hirata[a, *]**

**Authors' affiliations**

[a] Graduate School of Science, Kobe University, Kobe, Japan.

* Corresponding Author E-mail address: hirata@tiger.kobe-u.ac.jp

**Key Points**

Stratigraphic relationships between faults and impact craters were applied to determine the timing of the faulting.

The paucity of superposed craters implies that the faulting is 0.30–0.79 Ga.

This indicates that the faulting of the Wispy Terrain on Dione is a very recent event.

**Abstract**

The trailing hemisphere of Dione is characterized by the Wispy Terrain, where it exhibits a hemispheric-scale network of extensional tectonic faults superposed on the moon's cratered surface. The faults likely reflect past endogenic activity and Dione's interior thermal history. Although fresh exposures of pristine scarps indicate that the timing of the faulting is relatively recent, the absolute age of the faulting remains uncertain. To estimate the timing of the faulting, we investigated stratigraphic relationships between impact craters and faults. Using high-resolution images obtained by ISS cameras onboard the Cassini spacecraft, we investigated craters with diameters exceeding or equal to 10 km that coincide spatially with the faults, and classified the craters as crosscut craters or superposed craters. As a result, at least 82% of the craters were interpreted as clear examples of crosscut craters and 12% of the craters were interpreted to be candidates of superposed craters, although stratigraphic relationships are often ambiguous. The paucity of superposed craters and a predicted cratering rate indicate that the faulting of the Wispy Terrain is 0.30–0.79 Ga. If 12–18% of the craters are assumed to be superposed, the timing of the faulting could be in the range 0.30–0.79 Ga. However, it is possible that the faulting of the Wispy Terrain is still ongoing.

## 1. Introduction

Tectonic structures are prominent on Dione, and analyses of the faults can provide insight into its geologic history. Dione, with a radius of 561 km, is a mid-sized satellite in the Saturn system. Dione is Saturn's third densest moon, and its silicate mass fraction is estimated to be 50% [*Matson et al.*, 2009; *Thomas et al.*, 2007]. Radioactivity from the high amount of internal silicate material provided heat to Dione's interior and likely contributed to prolonged endogenic activity [*Consolmagno and Lewis*, 1978; *Hillier and Squyres*, 1991; *Moore*, 1984]. For example, tectonic fault systems in the Wispy Terrain (Fig. 1a) are suggested to be relatively young geologic features [*Kirchoff and Schenk*, 2015; *Stephan et al.*, 2010], although the timing of the faulting is still uncertain. To constrain the timing, we analyzed stratigraphic relationships between faults and craters.

The Wispy Terrain, defined as the area containing tectonic fault systems, mostly covers Dione's trailing hemisphere. A network of bright linear features on Dione's trailing hemisphere (Fig. 1a), called the wispy streaks or wispy markings, was first discovered during the Voyager flybys [*Smith et al.*, 1981; *Smith et al.*, 1982] and interpreted as bright exposures along troughs formed by either extensional tectonics [*Moore*, 1984; *Plescia and Boyce*, 1982] or cryovolcanism [*Plescia*, 1983; *Stevenson*, 1982]. Higher-resolution images, obtained by the ISS camera onboard the Cassini spacecraft (Fig. 1b), revealed that these bright linear features are sets of sub-parallel grabens with interstitial horsts in some regions [*Wagner et al.*, 2006]. The presence of these grabens and horsts are indicative of normal faulting induced by extensional tectonism [e.g., *Jaumann et al.*, 2009; *Stephan et al.*, 2010; *Wagner et al.*, 2006]. The fault scarp of Padua Chasmata has a slope of ~23°, which is shallower than expected from laboratory deformation experiments; so, modifications such as viscous relaxation have been suggested to explain such shallow slopes [*Beddingfield et al.*, 2015]. The existence of many polygonal impact craters in the Wispy Terrain implies that numerous subtle or nonvisible fractures extend across the surface of Dione [*Beddingfield et al.*, 2016].

The fault scarps of the Wispy Terrain are relatively fresh; high-resolution images (for example, Fig. 1c) indicate that slopes of the scarps show a paucity of landforms associated with mass wasting or erosion caused by impact cratering. This view received further support from *Stephan et al.* [2010], who found that the fault scarps expose clean $H_2O$ ice. This fresh exposure implies that formation of the fault scarps could have continued into a relatively recent time. However, such fresh exposure may alternatively be explained by recent small downslope movements. If so, the main faulting may be much older.

*Kirchoff and Schenk* [2015] estimated the age of the Wispy Terrain to be ~2.5 Ga based on crater density. Similarly, *Wagner et al.* [2006] estimated the age of the Wispy Terrain to be either older than 3.7 Ga or older than 1 Ga. However, these estimated ages are applicable to only the underlying surface and not applicable to the faulting itself [*Kirchoff and Schenk*, 2015; *Stephan et al.*, 2010], because these researchers did not evaluate whether each impact crater was formed after or before the faulting. As both of these studies proposed, the faults should be younger than the Wispy Terrain.

## 2. Data and Methods

We investigated the stratigraphic relationships between faults and craters to better constrain the fault ages. We used the methodology applied to faults on the Moon and Mercury [e.g. *Banks et al.*, 2015; *Byrne et al.*, 2014; *Watters et al.*, 2012]. Such studies commonly examine crosscut craters and superposed craters on fault scarps. They assume that a crater crosscut by scarps is formed before the faulting and that a crater superposed on scarps is formed after the faulting. We investigated impact craters that coincide spatially with the faults of the Wispy Terrain in addition to the superposition relationships between impact craters and faults.

### 2.1. Cassini Images

We used 95 images obtained by eight close encounters of the Cassini spacecraft (orbits B, 16, 50, 98, 137, 165, 214, and 217) (Table 1 and Fig. 2a). A complete list of the images is included as Supplementary Table S1. The fault systems of the Wispy Terrain are divided into six units: Carthage Fossae (CA), Clusium Fossae (CL), Palatine Chasmata (PL), Eurotas Chasmata (EU), Aurunca Chasmata (AU), and Padua Chasmata (PD) (Fig. 2b) [*Jaumann et al.*, 2009; *Roatsch et al.*, 2009]. The images cover almost all of the fault systems of the Wispy Terrain, except for the west end of PL, at a resolution of 25 m/px to 700 m/px (Fig. 2c, 2d). We focused on craters with diameters larger than 10 km ($D_c \geq$ 10km), which would be at least 20 pixels across the images used, to assure that the evaluation of the stratigraphic relationships was not strongly affected by image resolution, although classification of larger craters would be more reliable. We did not consider craters with diameters smaller than 10 km ($D_c <$ 10km). Note that $D_c$ stands for crater diameter.

### 2.2. Crater Classification

We classified all craters coinciding spatially with the faults into four types based on the intersection between crater and fault scarps (Table2, Fig. 3). A Type A crater is a crater in which fault scarps can be identified on the crater floor. A Type B crater has identifiable fault scarps on the rim crest and inner wall of the crater, but not on the floor. A Type C crater has fault scarps adjacent to the crater, but not on the floor or the inner wall of the crater, and a Type D crater is a crater that is unrelated to the faults (i.e., the crater is sufficiently far from the nearest faults). We defined Types A, B, and C as impact craters coinciding spatially with the faults. Although most of the craters on the Wispy Terrain are Type D craters, it is impossible to evaluate their superposition relationships, so we do not discuss Type D craters further.

Because the floor of a crater is generally filled by a breccia lens [*Grieve et al.*, 1989; *Melosh*, 1989], we considered Type A craters to be clear examples of craters formed before the faulting. On the other hand, the superposition of Type B craters could be uncertain for the following reasons. Many of the grabens on the Wispy Terrain are often faint (or not visible in available images) at topographic lows (Fig. 4), presumably because the depth of each fault is typically shallower than surface undulations. Following structural analyses of lunar grabens [*Golombek*, 1979; *McGill*, 1971] and assuming the width and the dip of the faults within a typical graben of the Wispy Terrain to be 5 km and 23° (the dip obtained by Beddingfield et al. [2015]), the base of the faulted layer is at the depth of 1 km, which is less than the depth of the craters. Based on this argument, most of the Type B craters are examples of craters formed before the faulting. However, Type B craters may be craters formed after the faulting because faults on an inner crater wall could be pre-existing faults exposed by the collapse of a transient crater wall. Type C craters are candidates for craters that formed after the faulting. However, some Type C craters may have formed before the faulting because even Type C craters could be alternatively explained by an ejecta blanket that sufficiently obscures faults near the crater rim.

Generally, in the case of the Moon or Mars, the extent of a continuous ejecta blanket is effective for stratigraphic classification because the thickness of the ejecta blanket is sufficient to obscure pre-existing grabens [e.g. *Kneissl et al.*, 2015; *Smith et al.*, 2009]. On the other hand, ejecta blankets of craters on Dione generally have insufficient thickness to obscure pre-existing faults presumably because the low gravity of Dione allows ejecta to travel much farther from the impact site. In fact, it is known that only a few craters have recognizable ejecta blankets on Dione [*Plescia*, 1983]. Therefore, to obtain a correct stratigraphic classification, we examined the rim crest and the floor of each crater and did not rely on ejecta blankets.

Note that the visibility of fault scarps also depends on the illumination or resolution of the images. For example, the three images in Figure 5 show the same impact crater in orbits 50, 137, and B (from left to right) projected at the same scale and the same projection. Orbits 50 and 137 did not image scarps on the crater floor or inner wall due to shadow and illumination. Therefore, a crater in the center of the figure would be classified as Type C if the classification was based on the image of orbit 50 or 137, but it would be Type A if based on the three images. Thus, a Type C crater could actually be a Type A or B crater, and a Type B crater could be a Type A. Because of the reasons mentioned in this subsection, this type of classification or stratigraphic interpretation is often ambiguous.

### 2.3. Identification and Measurements

Impact craters were identified by the existence of a circular, elliptical, or polygonal depression. The fault scarps were determined by linear slopes where bright material has been exposed, possibly due to normal faulting. The identifications were assessed by using the eight flyby images and the digital elevation model developed by *Gaskell* [2013] (Fig. 6). The flyby images provide different illumination of the Wispy Terrain and were useful for understanding the surface undulations, such as faults and craters. For example, the solar incidence during the flyby of orbit B brightened the east-facing slopes of the craters or scarps of PL and EU, whereas the incidence during the flybys of orbits 50 and 90 brightened the west-facing slopes of PL and EU. Most of the faults of the Wispy Terrain were covered by images from more than two flybys. Moreover, the digital elevation models of Dione with a spatial resolution of approximately 1.5 km, developed by *Gaskell* [2013], were used to improve the identification of scarps and crater depressions. The interpretations are organized in sketch maps (Supplementary Figs. S1-S3).

The Integrated Software for Imagers and Spectrometers (ISIS3), produced by the U.S. Geological Survey, was used to calibrate the Cassini images. To measure the diameters and the locations of impact craters, this investigation utilized the Java Mission-planning and Analysis for Remote Sensing (JMARS) program produced by Arizona State University. The craters were measured from the maps shown in Figure 2a.

## 3. Results

We identified 387 examples of craters coinciding spatially with the faults. The diameters, locations, and types of all the craters are shown in Supplementary Table S2. For craters ($D_c \geq 10$ km), 82% of the total are Type A, 6% are Type B, and 12% are Type C (Table 3).

Craters classified as Type C, with diameters larger than 25 km, are shown in Fig. 7. Although the rims and walls of the six craters are not cut by faults, there are small-scale faults adjacent to the craters but no faults on the opposite side of the craters. Thus, it is likely that the surrounding faults did not reach beyond the crater rim crests when the faults formed. Because most of the Type C craters with 25 km > $D_c \geq 10$ km are similar to these examples, craters formed after the faulting may be limited to only a small fraction of Type C craters. Complete illustrations for all craters (Types A, B, and C) are included in Supplementary Figures S4–S17. Note that some of the craters smaller than 10 km provide reliable examples of craters superposed on faults (Fig. 8), although we do not discuss these craters in detail.

## 4. Discussion

For age determination of the linear surface features, we followed the ideas developed by many authors [e.g., *Kneissl et al.*, 2015; *Smith et al.*, 2009; *Tanaka*, 1982; *Wichman and Schultz*, 1989]. These ideas assume that the area to use to obtain the crater number density is defined by the area enclosed by one crater radius away from the linear features (in this case, the faults) (Fig. 9). If the linear features are sparsely distributed and the length of the linear features ($L$) is sufficiently longer than either one crater radius ($r$) or the width of the linear features, the area ($A$) is simply determined by $A(r) = 2Lr$ .

However, because the faults of the Wispy Terrain are densely distributed, we needed to compute the value of $A(r)$ numerically. Based on the mapping of the fault systems (Supplementary Figs. S1–S3), the area enclosed by the radius $r$ [km] from the fault systems of the Wispy Terrain is approximated well by the equation

$$A(r) = 9370.79r + 557653 - 1180990/r \quad [\mathrm{km}^2]. \tag{1}$$

Following *Tanaka* [1982], the crater number density ($N_c$) can be extrapolated by the sum of the individual counts:

$$N_c = \sum_{i=1}^{k} \frac{1}{A(r_i)} \quad [\mathrm{km}^{-2}], \tag{2}$$

where $r_i$ is the radius of the $i$-th crater and $i = \{1, \ldots, k\}$ are defined by all craters larger than a given size.

The heliocentric cratering rate was estimated theoretically by *Zahnle et al.* [2003]. Because of the uncertainty of the cratering rate in the outer solar system, they defined two cases: Case A, which is extrapolated from the cratering rate in the Jovian system, and Case B, which is extrapolated from the cratering rate for Triton. Thus, the theoretical cratering rate ($R_A$ and $R_B$) for craters larger than a given crater diameter ($D_c$)

on Dione is given by

$$R_A = 3.0 \times 10^{-14} \left(\frac{D_c}{10}\right)^{-1.277} \quad [\text{km}^{-2}\ \text{year}^{-1}] \text{ for Case A, or} \qquad (3)$$

$$R_B = 1.9 \times 10^{-13} \left(\frac{D_c}{10}\right)^{-2.171} \quad [\text{km}^{-2}\ \text{year}^{-1}] \text{ for Case B.} \qquad (4)$$

We dated the fault systems using Eqs. (2) to (4).

If we assume that Type A craters are crosscut craters and that Types B and C craters are superposed craters, the crater number density for superposed craters ($D_c \geq 10$ km) is 149.28 craters/$10^6$ km$^2$. Thus, the age of the faulting is estimated to be 5.0 Gyr in Case A and 0.79 Gyr in Case B. If we assume that Types A and B are crosscut craters and that Type C is superposed craters, the crater number density for the superposed craters ($D_c \geq 10$ km) is 101.67 craters/$10^6$ km$^2$. Thus, the age of the faulting is estimated to be 3.4 Gyr in Case A and 0.54 Gyr in Case B. If the timing of the faulting is estimated to be 18% (Types B and C) or 12% (only Type C) of 2.5 Ga, it would be 0.45 Ga or 0.30 Ga, respectively. Considering the existence of Type A craters (82% of the total) and the age of the Wispy Terrain (~2.5 Ga) determined by *Kirchoff and Schenk* [2015], the age based on Case A is unlikely. Overall, the timing of the faulting is estimated to be between 0.30 and 0.79 Ga. Those estimated ages are organized into Table 4. Considering that most Type C craters are not very clear examples of superposed craters, the timing of the faulting may be much younger than the age obtained. Therefore, the possibility that the faulting is an ongoing event cannot be ruled out.

Note that it is unlikely that the ages of these faults are the same, and each unit of the fault systems could have a variant age. In other word, the faulting may occur as multiple events, rather than a single event. Furthermore, the possibility of faults being reactivated cannot be ruled out, and therefore, craters superposed on faults at its formation may be cut by the reactivation of faults. If so, some of crosscut craters would reflect the timing of the reactivation instead. Therefore, the timing of the faulting obtained could be said to be the averaged value of all faults. In any case, the timing of the faulting of the Wispy Terrain is quite recent, possibly even ongoing. However, the stratigraphic relationships are ambiguous, as discussed in section 2.2; therefore, this method involves uncertainty.

Some of the crosscut craters in this study would be reclassified as superposed craters by other researchers. If 5% of the Type A craters were reclassified as superposed craters, the age would be 0.18 Gyr older than the original dating. Additional work on the stratigraphic relationship between the dark terrain and the faults of the Wispy Terrain may provide a more reliable age for the faulting. It is known that the faults of the Wispy

Terrain are superposed on the dark surface of the trailing hemisphere [*Stephan et al.*, 2010], and the dark surface is considered to be the result of the accumulation of dark materials during the last 1–50 Ma [*Hirata and Miyamoto*, 2016]. Interestingly, this age is consistent with this work.

## 5. Conclusion

More than 387 impact craters ($D_c \geq 10$ km) coinciding spatially with the fault system of the Wispy Terrain were identified and classified as crosscut craters or superposed craters based on geological investigation. As a result, we interpreted that at least 82% of the total craters examined are reliable examples of crosscut craters and that the rest are candidates of superposed craters. Based on the theoretical cratering rate, the fault ages are between 0.30 and 0.79 Ga. This indicates that the faulting is a very recent event. On the other hand, stratigraphic relations are often ambiguous, and the detailed timing and accuracy of this type of study remain unclear. For example, if 5% of the Type A craters were reclassified as superposed crates, the age would be 0.18 Gyr older than the original dating. Further studies, such as studies on the stratigraphic relationship between the dark surface of the trailing hemisphere and the faults of the Wispy Terrain, may provide a more reliable timing for the faulting.

## Figures

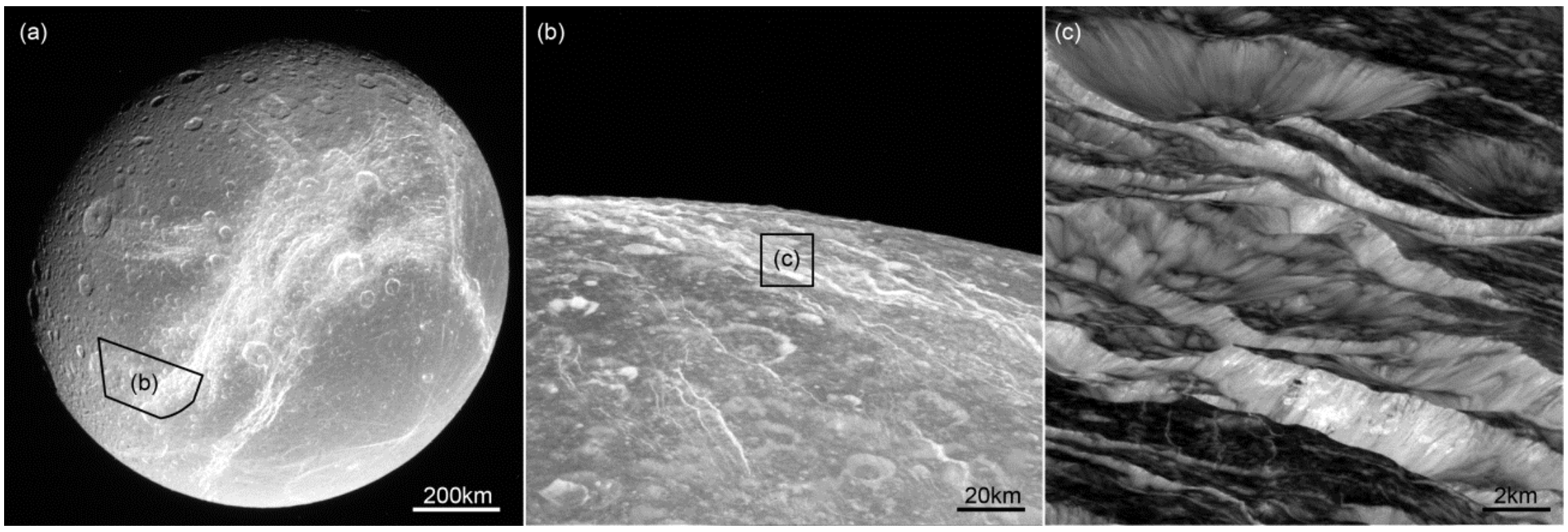

**Figure 1.** (a) Dione's trailing hemisphere (N1532405095). (b) A close-up view of the fault scarps of the Wispy Terrain (W1649311515). (c) One of the highest-resolution images of the faults (N1649311515; 14 m/px). Insets outline the locations of Figure 1b and 1c.

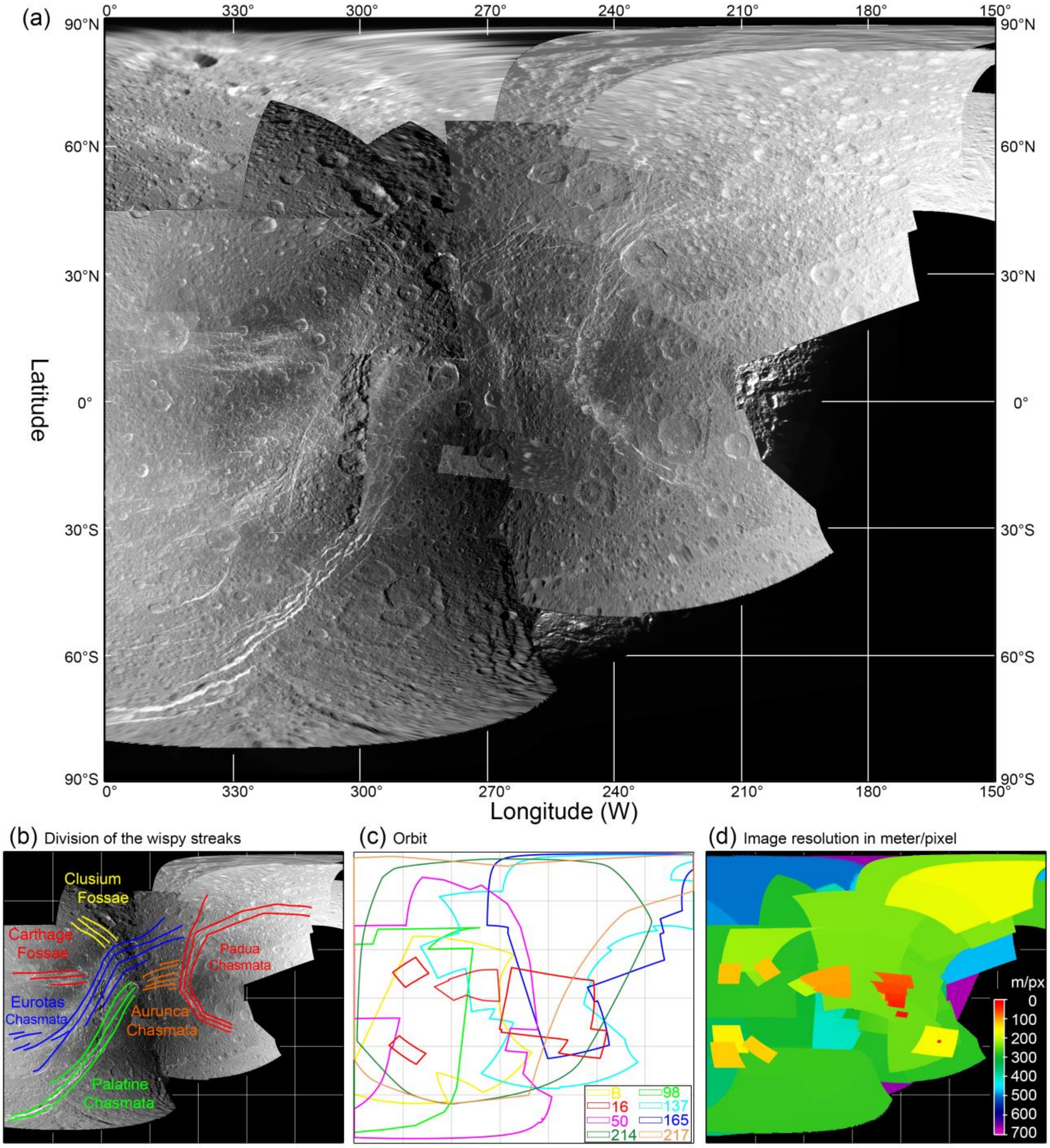

**Figure 2.** (a) Hemispheric mosaic of the Wispy Terrain in simple cylindrical projection. (b) Nomenclature for each division of the fault system. (c) The regions imaged in each orbit. (d) The image resolution for each location in m/px. Red indicates higher resolution, and blue indicates lower resolution. The base maps of (b), (c), and (d) are equivalent to the map in (a).

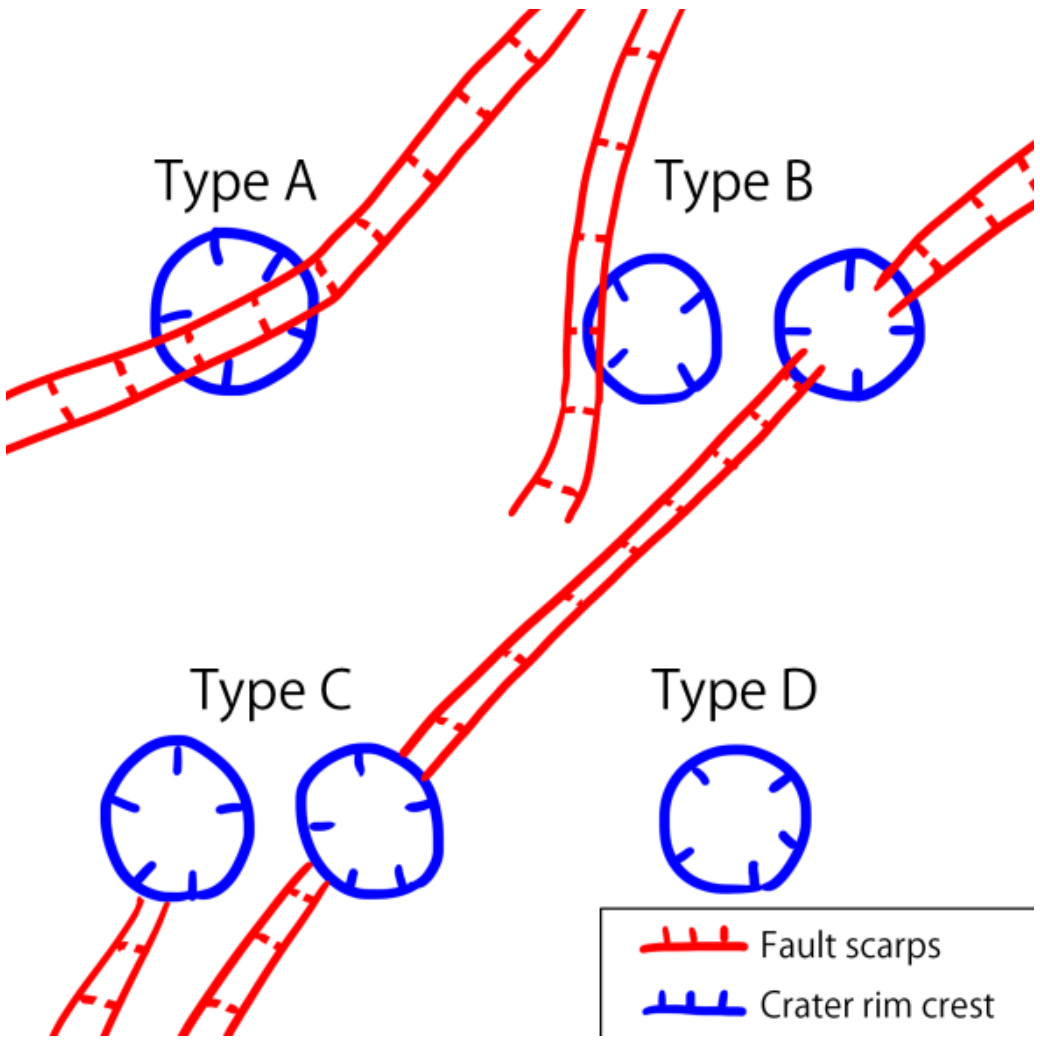

**Figure 3.** The definition of the classification of this work. A Type A crater is a crater with its floor cut by faults, a Type B crater is a crater with its inner wall but not its floor cut by faults, a Type C crater is a crater with an inner wall and floor not cut by faults, and a Type D crater is a crater sufficiently far from the nearest faults.

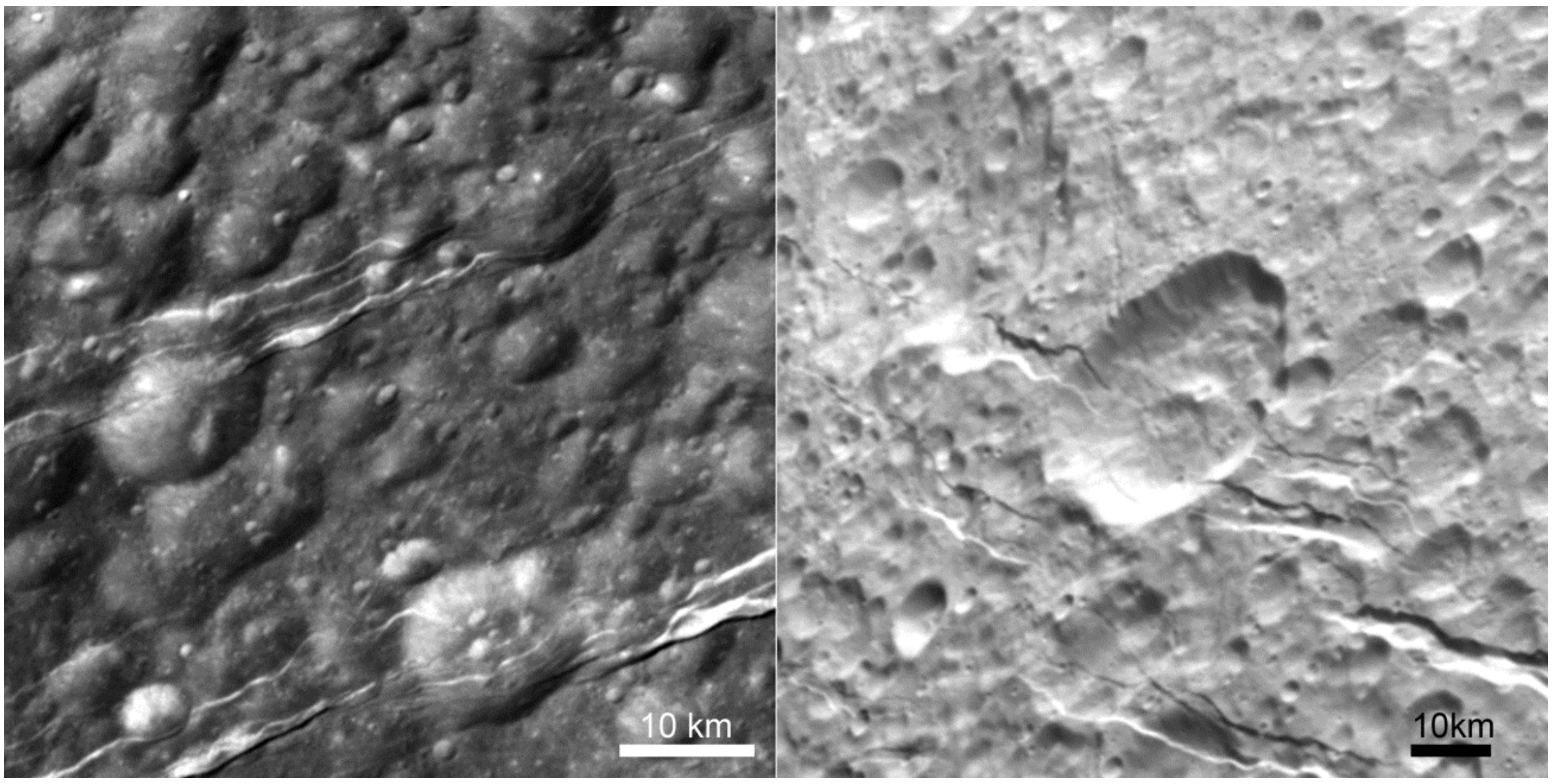

**Figure 4.** Faults scarps of the Wispy Terrain. Many of the grabens on the Wispy Terrain are faint or not visible in topographic lows (N1649313734 at left and N1662201249 at right).

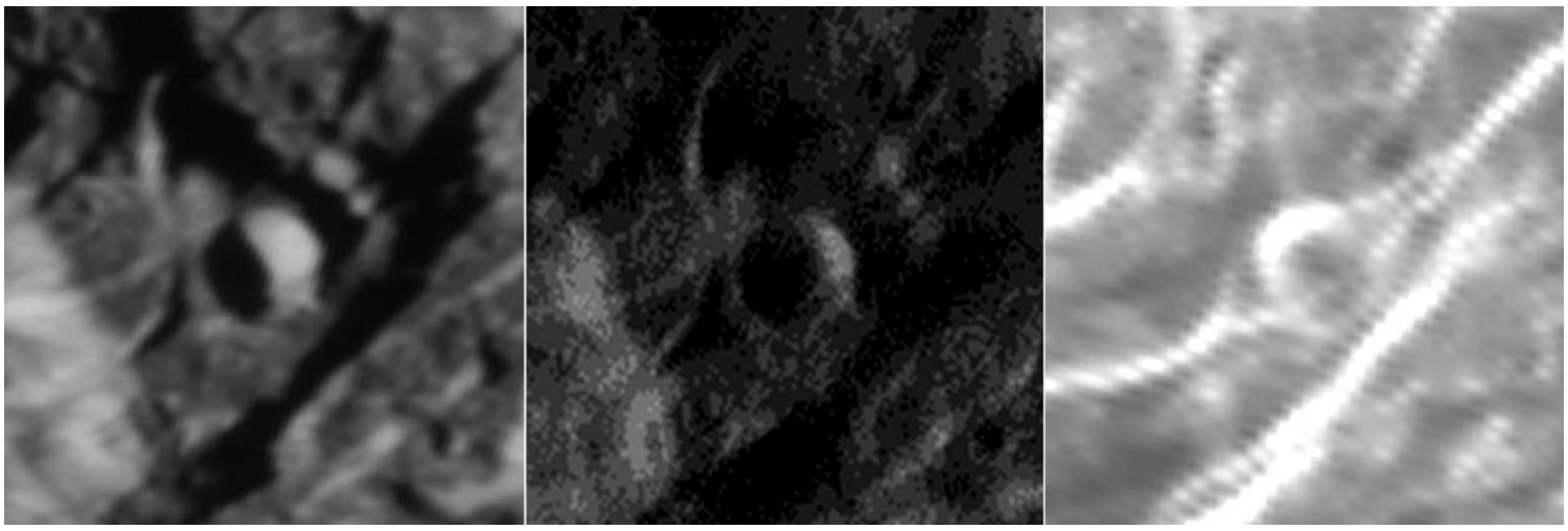

**Figure 5.** Example comparison among images from orbits 50, 137, and B (from left to right, respectively). The three images show the same impact crater (with a diameter of 7.61 km) centered at 13.2°N, 284.8°W and are shown at the same scale.

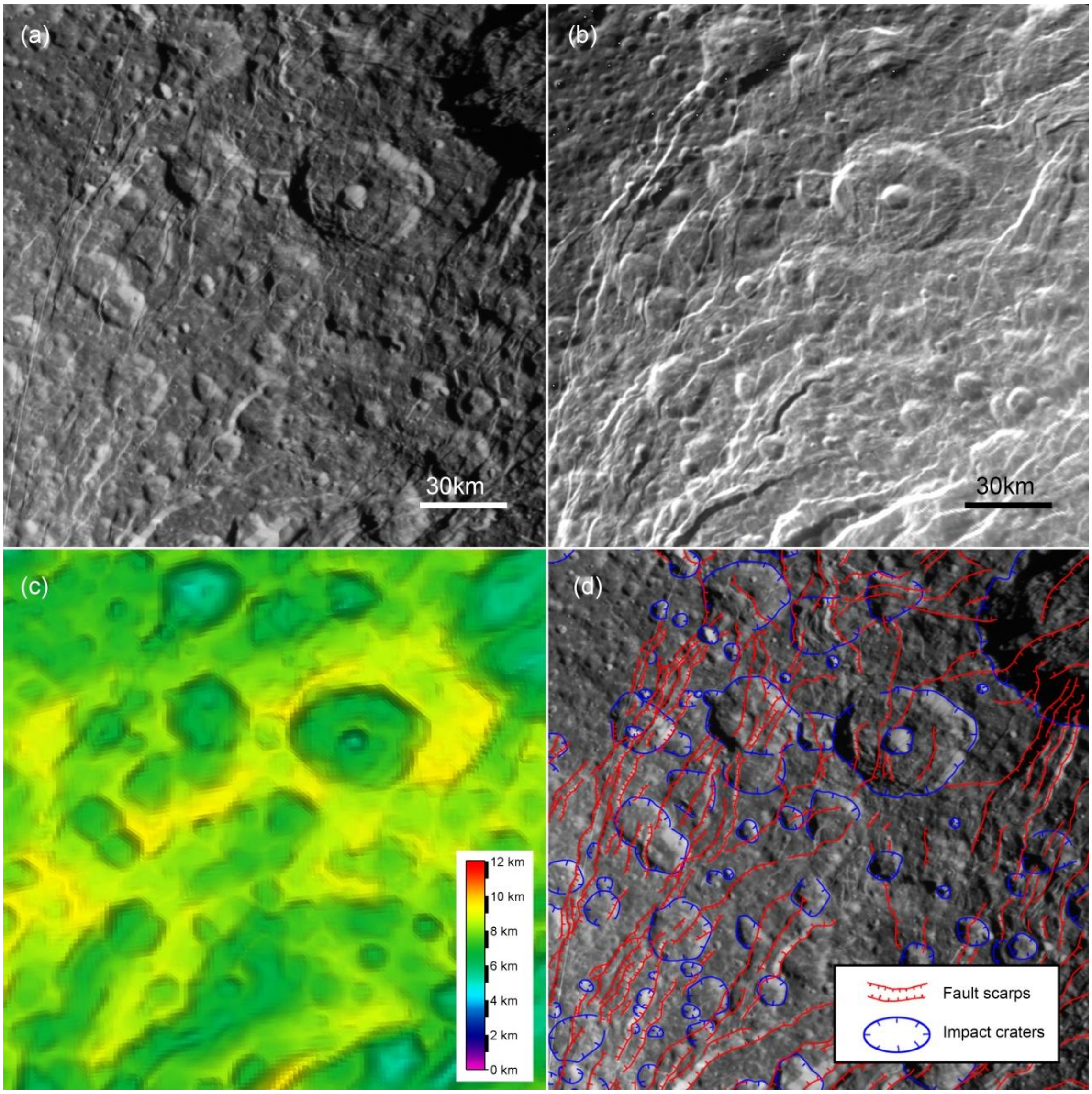

**Figure 6.** Example of the identification of the impact craters and the fault scarps of Eurotas Chasmata. The base map is centered at 292.1°W and 22.3°N, and its projection is simple cylinder. (a) Images in orbit 50. (b) Images in orbit B. (c) Topography from *Gaskell* [2013]. (d) Sketch map of the craters (blue lines) and the fault scarps (red lines) based on the investigation. The hachures point downscarp, and the lines without hachures indicate that the downscarp direction is uncertain.

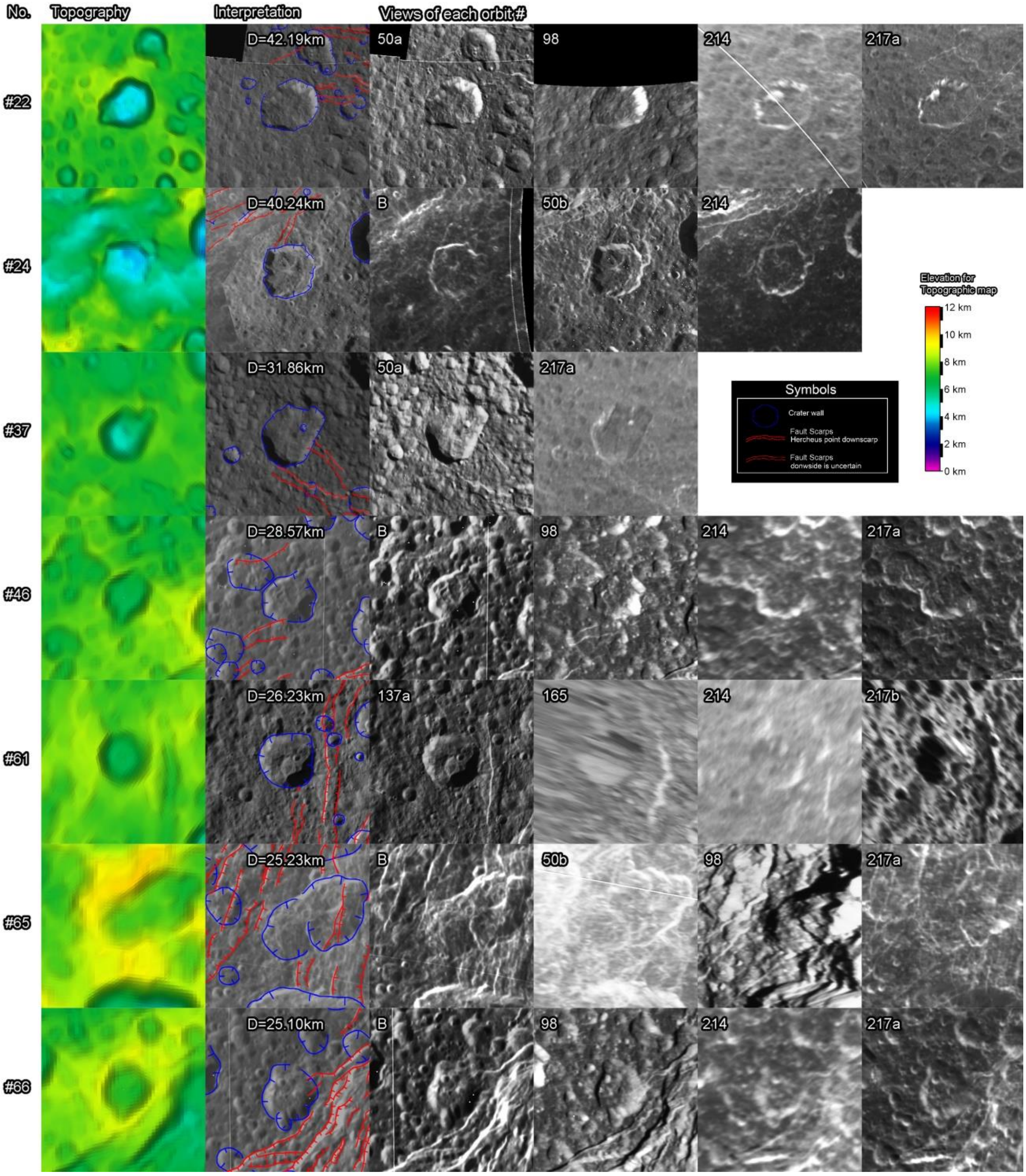

**Figure 7.** Type C craters with diameter larger than 25 km. Images shown in the same line are of the same crater projected in the same geometry; azimuthal equidistant projection centered at the crater (each crater's longitude and latitude are shown in Table

S2). The ground scale is set so that each crater diameter is one-third of the width of each image. The numbers at the left correspond to the crater numbers in Table S2. The left end column shows the topography (its elevation scale is shown in upper right of this figure), the second column from the left shows the interpretation of the faults and the craters in this work (its symbol is shown in the upper right), and the third and subsequent columns from the left to the right are mosaics merged from images obtained by the Cassini flybys. The other craters are shown in Supplementary Figs. S4–S17.

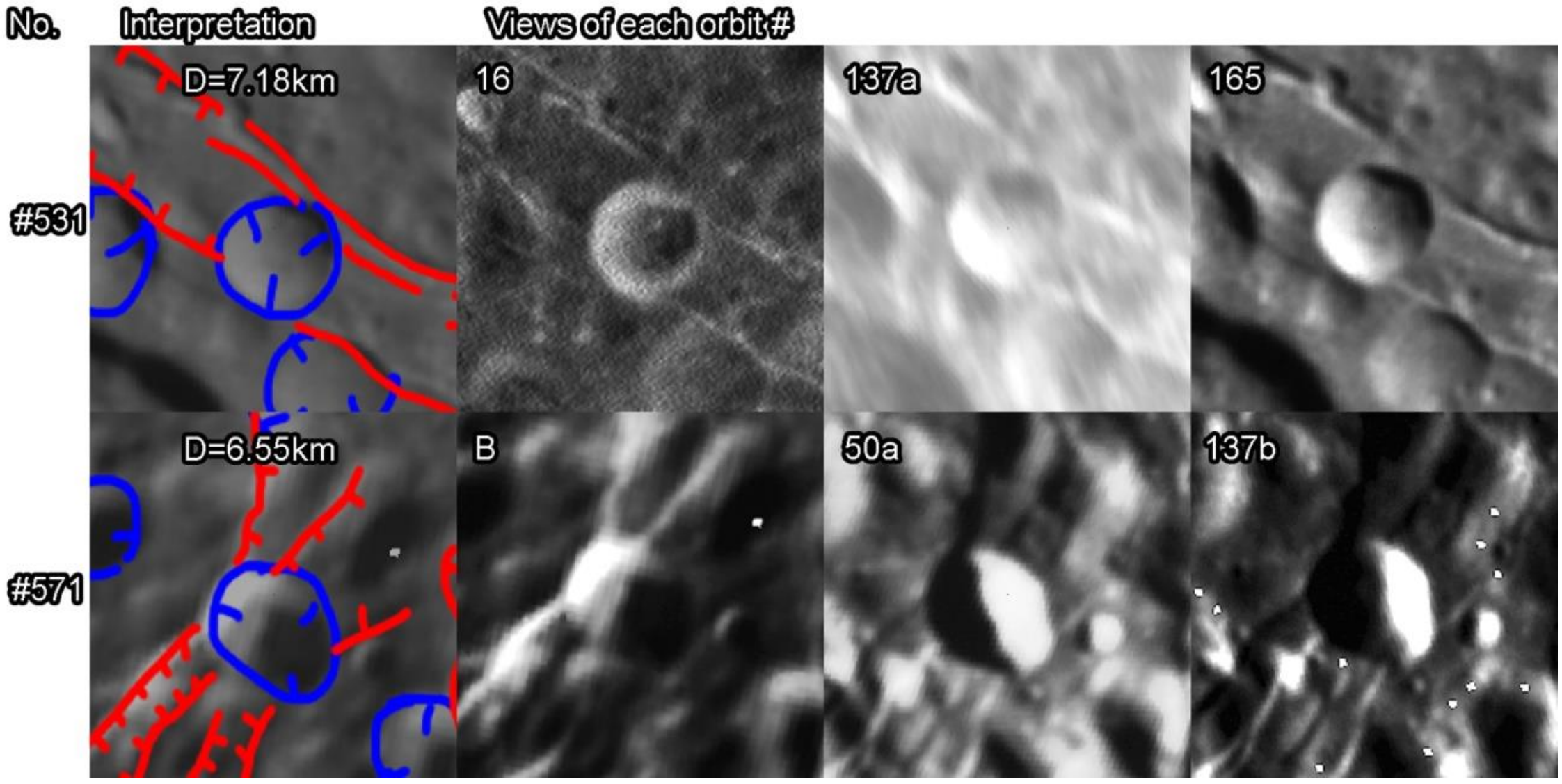

**Figure 8.** Candidates of superposed craters on faults. The left end column shows the interpretation of faults and craters in this work and the second and subsequent columns from left to right are mosaics merged from images obtained by the Cassini flybys.

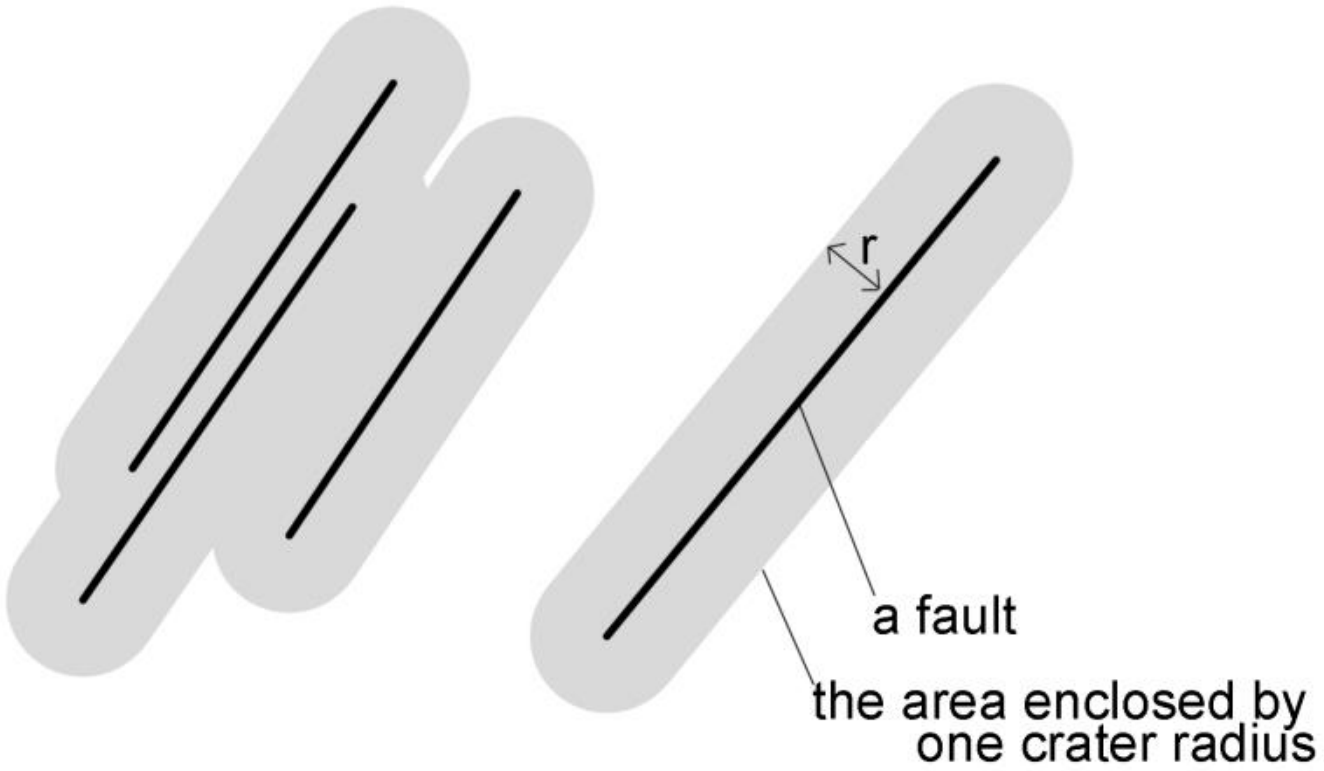

**Figure 9.** The concept of the definition of area to obtain crater density. The gray regions enclosed by a given radius ($r$) from a fault are equivalent to $A(r)$ in Eq. 1.

**Table 1 The 8 close flyby orbits used in this work.**

| Orbit | Date | Number of images | Resolution (m/px) |
|---|---|---|---|
| **B** | Dec. 15, 2004 | 4 | 432 |
| **16** | Oct. 12, 2005 | 19 | 25-360 |
| **50** | Sep. 30, 2007 | 13 | 272-415 |
| **98** | Apr. 7, 2010 | 10 | 115-353 |
| **137** | Sep. 4, 2010 | 17 | 235-304 |
| **165** | May 3, 2012 | 9 | 170-474 |
| **214** | Apr. 11, 2015 | 6 | 660 |
| **217** | June 16, 2015 | 13 | 437-522 |

**Table 2 Crater classification and interpretation of this work.**

| Crater type | Definition | Superposition |
|---|---|---|
| Type A | A crater in which fault scarps can be identified on the crater floor | before the faulting |
| Type B | A crater which has identifiable fault scarps on the rim crest and inner wall of the crater, but not on the floor | before or after the faulting |
| Type C | A crater which has fault scarps adjacent to the crater, but not on the floor or the inner wall of the crater | after the faulting |
| Type D | A crater unrelated to the faults (i.e., the crater is sufficiently far from the nearest faults) | uncertain |

**Table 3 The Number of each type of craters.**

| | Type A | Type B | Type C | Total |
|---|---|---|---|---|
| $D_c \geq 50km$ | 13 | 1 | 0 | 14 |
| $D_c \geq 30km$ | 37 | 3 | 3 | 43 |
| $D_c \geq 20km$ | 100 | 4 | 12 | 116 |
| $D_c \geq 10km$ | 316 | 23 | 48 | 387 |

**Table 4 The estimated age of the faulting.**

| Assumption | Case A | Case B | Other * |
|---|---|---|---|
| If Type A is crosscut while Type B and C are superposed | 5.0 Gyr | 0.79 Gyr | 0.45 Gyr |
| If Type A and B are crosscut while Type C is superposed | 3.4 Gyr | 0.54 Gyr | 0.30 Gyr |

*based on the age of the Wispy Terrain (2.5 Gyr) and the fraction of crosscut craters out of the total craters.

## Acknowledgements

The author thanks Dr. Chloe Beddingfield, Dr. Michelle Kirchoff, and Dr. Noah Hammond for their helpful comments and suggestions, which significantly improved this work. This work was supported by a Grant-in-Aid for JSPS Fellows. The image data (http://pds.nasa.gov) and the shape model of Dione (http://sbn.psi.edu/pds/asteroid/) are freely available via NASA's Planetary Data System. The softwares ISIS3 (http://isis.astrogeology.usgs.gov/) and JMARS (https://jmars.asu.edu/) were used in this work.

**Supporting information of this manuscript**

**Introduction**

This file includes Tables S1, S2, and the descriptions for supplementary figures and tables. Figures S4-S17 include 700 craters coinciding spatially with faults. Craters down to 5 km in diameter are listed in order to obtain completeness at 10 km.

**Figure Captions**

**Figure S1** Sketch maps of the craters (blue lines) and the fault scarps (red lines) based upon the investigation, which are shown in simple cylindrical projection. The region higher latitude than N45° is shown in Fig S2. The region lower latitude than S45° is shown in Fig S3.

**Figure S2** Sketch maps of the craters (blue lines) and the fault scarps (red lines) in north polar stereographic projection.

**Figure S3** Sketch maps of the craters (blue lines) and the fault scarps (red lines) in south polar stereographic projection.

**Figure S4** Craters from No. 1 to No. 50. Images shown in the same line are of the same crater projected in the same geometry; azimuthal equidistant projection centered at the crater (each crater's longitude and latitude are shown in Table S2). The ground scale is set so that each crater diameter is one-third of the width of each image. The numbers at the left correspond to the crater numbers in Table S2. The left end column shows the topography (its elevation scale is shown in upper right of this figure), the second column from the left shows the interpretation of the faults and the craters in this work (its symbol is shown in the upper right), and the third and subsequent columns from the left to the right are mosaics merged from images obtained by the Cassini flybys. This work developed 11 mosaics (B, 16, 50a, 50b, 98, 137a, 137b, 165, 214, 217a, and 217b). For example, the mosaic 165 was made from the 9 images whose most right column of Table S1 noted 165. In the case of images during the 3 flybys (orbit #50, #137, and #217), the images obtained are categorized into two types based on the different in resolution or illumination, and thus, this work made two distinct mosaics for the 3 flybys (for example, 50a and 50b).

**Figure S5** Craters from No. 51 to No. 100.

**Figure S6** Craters from No. 101 to No. 150.

**Figure S7** Craters from No. 151 to No. 200.

**Figure S8** Craters from No. 201 to No. 250.

**Figure S9** Craters from No. 251 to No. 300.

**Figure S10** Craters from No. 301 to No. 350.

**Figure S11** Craters from No. 351 to No. 400.

**Figure S12** Craters from No. 401 to No. 450.

**Figure S13** Craters from No. 451 to No. 500.

**Figure S14** Craters from No. 501 to No. 550.

**Figure S15** Craters from No. 551 to No. 600.

**Figure S16** Craters from No. 601 to No. 650.

**Figure S17** Craters from No. 651 to No. 700.

**Tables**

Table S1 shows a complete list of the images used in this work. These images are freely available from NASA's Planetary Data System. Note that the three flybys in orbits B, 50, and 98 provide images of the western part of the Wispy Terrain (CA, CL, PL, and EU), the two flybys in orbits 137 and 165 provide images of the eastern part of the Wispy Terrain (AU and PD), the flyby in orbit 16 provides the highest-resolution views of limited small areas of the Wispy Terrain, and the two flybys in orbit #214 and #217 provide entire of the Wispy Terrain at high sun. Table S2 shows the location and diameter of craters coinciding spatially with the faults of the Wispy Terrain.

**Table S1 Images used in this work**

| Image | Lat. | Lon. (W) | Resolution(m) | Mosaic* |
|---|---|---|---|---|
| orbit #B (December 15, 2004) | | | | |
| N1481766784_1 | 14.03 | 329.73 | 432.87 | B |
| N1481766908_1 | -21.17 | 311.18 | 433.63 | B |
| N1481767088_1 | -11.12 | 298.48 | 433.15 | B |
| N1481767211_2 | 11.87 | 297.77 | 432.51 | B |
| orbit #16 (October 12, 2005) | | | | |
| N1507744755_2 | 10.48 | 240.77 | 68.82 | 16 |
| N1507744823_3 | 9.28 | 240.10 | 65.06 | 16 |
| N1507744919_2 | 5.49 | 244.20 | 60.03 | 16 |
| N1507744987_2 | 4.44 | 243.33 | 56.26 | 16 |
| N1507745052_2 | 0.87 | 243.54 | 105.44 | 16 |
| N1507745166_2 | -0.52 | 242.58 | 46.46 | 16 |
| N1507745229_2 | -3.43 | 242.26 | 86.00 | 16 |
| N1507745279_2 | -3.92 | 242.00 | 40.28 | 16 |
| N1507745347_2 | -4.56 | 241.58 | 36.57 | 16 |
| N1507745423_2 | -10.74 | 238.92 | 32.29 | 16 |
| N1507745456_2 | -10.74 | 238.94 | 30.56 | 16 |
| N1507745557_2 | -10.82 | 238.96 | 25.08 | 16 |
| N1507745708_2 | -27.46 | 215.92 | 32.21 | 16 |
| N1507747826_2 | 4.31 | 283.31 | 97.12 | 16 |
| N1507747896_2 | 11.92 | 288.72 | 100.64 | 16 |
| N1507747961_2 | 8.16 | 295.24 | 207.67 | 16 |
| N1507748010_2 | 8.51 | 293.01 | 106.62 | 16 |
| N1507748185_2 | 15.18 | 326.11 | 229.65 | 16 |
| N1507748228_2 | 15.62 | 325.30 | 117.20 | 16 |
| N1507748334_2 | -33.95 | 329.54 | 123.17 | 16 |
| W1507745347_2 | -6.05 | 230.61 | 360.08 | 16 |
| W1507745557_2 | -11.34 | 235.03 | 248.74 | 16 |
| W1507745708_2 | -27.55 | 216.21 | 161.08 | 16 |
| orbit #50 (September 30, 2007) | | | | |
| N1569814805_1 | -46.47 | 289.48 | 415.40 | 50b |
| N1569815121_1 | -21.59 | 289.41 | 407.07 | 50b |
| N1569815285_1 | 2.40 | 256.55 | 402.07 | 50b |
| N1569815436_1 | 10.46 | 285.83 | 399.22 | 50b |
| N1569815593_1 | 29.06 | 271.48 | 395.37 | 50b |
| N1569826692_3 | -45.54 | 282.24 | 271.77 | 50a |
| N1569826794_3 | -49.86 | 316.91 | 272.43 | 50a |
| N1569827462_1 | 2.24 | 319.63 | 278.55 | 50a |
| N1569827571_1 | 2.02 | 316.09 | 280.04 | 50a |
| N1569827692_1 | 27.43 | 304.74 | 281.90 | 50a |
| N1569827799_1 | 27.15 | 316.77 | 283.35 | 50a |
| N1569827906_1 | 24.87 | 283.01 | 285.78 | 50a |
| N1569828025_1 | 50.81 | 299.91 | 287.63 | 50a |
| orbit #98 (April 7, 2010) | | | | |
| N1649313601_1 | 14.46 | 345.91 | 115.34 | 98 |
| N1649313734_1 | -35.50 | 348.40 | 122.45 | 98 |
| N1649314237_1 | -25.72 | 346.68 | 147.52 | 98 |
| N1649315659_1 | 28.52 | 333.13 | 219.95 | 98 |
| N1649317009_1 | -40.99 | 318.55 | 289.71 | 98 |
| N1649317172_1 | -15.60 | 318.07 | 297.74 | 98 |
| N1649317349_1 | 12.03 | 319.48 | 306.79 | 98 |
| N1649317889_1 | 14.24 | 212.26 | 333.58 | 98 |
| N1649318068_1 | -17.77 | 221.09 | 342.95 | 98 |
| N1649318247_1 | -47.99 | 245.66 | 353.30 | 98 |
| orbit #137 (September 4, 2010) | | | | |
| N1662196086_1 | 31.78 | 298.98 | 292.97 | 137b |
| N1662198548_1 | 8.45 | 276.66 | 238.08 | 137a, 137b |
| N1662198718_1 | 31.35 | 277.99 | 236.16 | 137a, 137b |
| N1662198888_1 | 52.36 | 278.22 | 235.12 | 137a, 137b |
| N1662199639_1 | 69.09 | 220.21 | 235.37 | 137a |
| N1662199809_1 | 49.63 | 246.41 | 235.69 | 137a |
| N1662199979_1 | 28.88 | 251.87 | 236.91 | 137a |
| N1662200149_1 | 5.51 | 254.18 | 239.02 | 137a |
| N1662200504_1 | -18.46 | 234.30 | 244.65 | 137a |
| N1662200736_2 | 5.24 | 231.38 | 247.36 | 137a |
| N1662200906_1 | 27.49 | 224.42 | 250.02 | 137a |
| N1662201078_1 | 46.36 | 207.56 | 253.67 | 137a |
| N1662201249_1 | 57.40 | 165.60 | 258.35 | 137a |
| N1662202012_1 | -21.42 | 219.73 | 278.41 | 137a |
| N1662202200_1 | -22.06 | 247.80 | 283.95 | 137a |
| N1662202465_1 | 2.36 | 252.03 | 291.45 | 137a |
| N1662202839_1 | 32.64 | 238.82 | 303.58 | 137a |
| orbit #165 (May 3, 2012) | | | | |
| N1714687039_1 | 60.87 | 178.54 | 333.80 | 165 |
| N1714687090_1 | 61.63 | 180.17 | 169.47 | 165 |
| N1714687261_1 | 53.42 | 207.77 | 177.79 | 165 |
| N1714688407_1 | 50.79 | 194.42 | 234.27 | 165 |
| N1714688580_1 | 39.12 | 220.97 | 243.10 | 165 |
| N1714688749_1 | 21.91 | 235.91 | 251.85 | 165 |
| N1714688907_1 | 0.95 | 232.54 | 259.51 | 165 |
| N1714689063_1 | -19.99 | 225.95 | 267.21 | 165 |
| N1714693066_1 | 36.95 | 205.54 | 474.67 | 165 |
| orbit #214 (April 11, 2015) | | | | |
| N1807488286_1 | -22.21 | 303.35 | 662.35 | 214 |
| N1807488394_1 | 23.57 | 303.88 | 662.01 | 214 |
| N1807489428_1 | -25.33 | 268.90 | 660.27 | 214 |
| N1807489536_1 | 27.21 | 268.89 | 660.30 | 214 |
| N1807491456_1 | -22.18 | 232.07 | 667.08 | 214 |
| N1807491547_1 | 23.90 | 231.20 | 667.67 | 214 |
| orbit #217 (June 16, 2015) | | | | |
| N1813168874_1 | 0.06 | 280.43 | 521.58 | 217a |
| N1813169097_1 | 28.33 | 301.69 | 510.43 | 217a |
| N1813169317_1 | 54.33 | 282.38 | 500.35 | 217a |
| N1813169555_1 | 29.37 | 127.19 | 488.53 | 217a |
| N1813169775_1 | 1.41 | 310.94 | 477.20 | 217a |
| N1813169995_1 | -26.57 | 305.22 | 467.62 | 217a |
| N1813170233_3 | -50.35 | 273.53 | 456.80 | 217a |
| N1813170453_1 | -24.79 | 120.35 | 445.99 | 217a |
| N1813170635_1 | 2.07 | 31.80 | 437.88 | 217a |
| N1813190432_1 | -25.02 | 230.79 | 444.64 | 217b |
| N1813190623_1 | 3.01 | 239.54 | 453.61 | 217b |
| N1813190854_1 | 31.51 | 217.93 | 464.04 | 217b |
| N1813191020_1 | 49.23 | 181.43 | 472.15 | 217b |

*Mosaic number shown in Figure S4-S17.

**Table S2 Craters identified in this work**

| Crater No. | Lat. | Lon. (W) | Diameter(km) | Type |
|---|---|---|---|---|
| #1 | 52.20 | 202.68 | 129.05 | A |
| #2 | 50.16 | 243.48 | 100.96 | B |
| #3 | −21.06 | 344.71 | 97.39 | A |
| #4 | 33.34 | 233.04 | 89.32 | A |
| #5 | 15.58 | 345.31 | 89.10 | A |
| #6 | 30.01 | 280.46 | 83.03 | A |
| #7 | 5.30 | 280.02 | 77.41 | A |
| #8 | −2.42 | 302.77 | 75.20 | A |
| #9 | −14.88 | 301.89 | 66.45 | A |
| #10 | −32.22 | 332.84 | 66.03 | A |
| #11 | −3.29 | 241.29 | 62.17 | A |
| #12 | −10.15 | 237.40 | 59.75 | A |
| #13 | 25.13 | 272.20 | 59.61 | A |
| #14 | 8.82 | 310.04 | 50.41 | A |
| #15 | −15.21 | 287.33 | 49.84 | A |
| #16 | 6.24 | 310.28 | 49.77 | A |
| #17 | −12.88 | 231.55 | 47.14 | A |
| #18 | 39.35 | 246.33 | 46.76 | A |
| #19 | −25.67 | 215.88 | 44.52 | A |
| #20 | 25.24 | 288.43 | 43.74 | A |
| #21 | 14.00 | 278.20 | 43.55 | A |
| #22 | 41.14 | 322.86 | 42.19 | C |
| #23 | 48.59 | 163.95 | 41.54 | A |
| #24 | −2.19 | 275.49 | 40.24 | C |
| #25 | 22.07 | 267.94 | 39.33 | A |
| #26 | −39.91 | 317.36 | 38.41 | A |
| #27 | 41.06 | 310.83 | 38.17 | A |
| #28 | 51.17 | 174.04 | 37.91 | A |
| #29 | 24.75 | 236.64 | 36.61 | B |
| #30 | −12.42 | 297.22 | 36.15 | A |
| #31 | −41.74 | 314.12 | 35.86 | A |
| #32 | 18.66 | 249.40 | 33.80 | A |
| #33 | −68.21 | 333.96 | 33.74 | A |
| #34 | −38.02 | 313.64 | 33.40 | A |
| #35 | 25.98 | 294.38 | 32.71 | A |
| #36 | 1.34 | 255.72 | 32.51 | A |
| #37 | 52.57 | 316.44 | 31.86 | C |
| #38 | −8.13 | 301.00 | 31.70 | A |
| #39 | 4.96 | 306.93 | 31.54 | A |
| #40 | 17.42 | 313.73 | 31.35 | A |
| #41 | 48.54 | 211.66 | 31.30 | A |
| #42 | 46.08 | 243.99 | 30.59 | B |
| #43 | 51.25 | 192.10 | 30.14 | A |
| #44 | −14.79 | 290.35 | 29.57 | A |
| #45 | 39.15 | 289.82 | 29.50 | A |
| #46 | −14.23 | 324.45 | 28.57 | C |
| #47 | 22.42 | 263.85 | 28.56 | A |
| #48 | −20.96 | 334.65 | 28.52 | A |
| #49 | 45.80 | 319.93 | 28.43 | A |
| #50 | 61.65 | 224.77 | 28.41 | A |
| #51 | 41.97 | 209.14 | 28.31 | A |
| #52 | 26.36 | 242.58 | 27.62 | A |
| #53 | 12.47 | 286.92 | 27.58 | A |
| #54 | 14.37 | 309.61 | 27.57 | A |
| #55 | 8.55 | 304.20 | 27.32 | A |
| #56 | 2.01 | 299.23 | 26.80 | A |
| #57 | 30.83 | 294.78 | 26.77 | A |
| #58 | 3.08 | 249.46 | 26.62 | A |
| #59 | 0.42 | 306.65 | 26.30 | A |
| #60 | −9.19 | 296.45 | 26.28 | A |
| #61 | 61.02 | 239.72 | 26.23 | C |
| #62 | 39.85 | 219.29 | 26.18 | A |
| #63 | 16.17 | 327.56 | 25.80 | A |
| #64 | 56.57 | 178.96 | 25.80 | A |
| #65 | −8.96 | 303.05 | 25.23 | C |
| #66 | −15.33 | 319.64 | 25.10 | C |
| #67 | −25.25 | 328.96 | 24.95 | A |
| #68 | −22.57 | 354.67 | 24.91 | A |
| #69 | −26.75 | 322.00 | 24.49 | A |
| #70 | 33.64 | 248.64 | 24.36 | A |
| #71 | −20.00 | 332.31 | 23.96 | A |
| #72 | 17.54 | 297.22 | 23.89 | A |
| #73 | −26.40 | 335.55 | 23.76 | A |
| #74 | 38.16 | 241.45 | 23.75 | A |
| #75 | 4.67 | 246.83 | 23.71 | A |
| #76 | −2.35 | 246.18 | 23.61 | A |
| #77 | 54.98 | 175.14 | 23.41 | A |
| #78 | −16.73 | 224.19 | 23.31 | A |
| #79 | 36.46 | 290.85 | 23.26 | A |
| #80 | 5.52 | 265.27 | 23.22 | C |
| #81 | −56.85 | 347.02 | 23.20 | C |
| #82 | −49.44 | 348.01 | 23.15 | C |
| #83 | 66.36 | 208.06 | 23.04 | A |
| #84 | −12.99 | 308.28 | 22.98 | A |
| #85 | 42.66 | 259.00 | 22.82 | A |
| #86 | 23.16 | 260.94 | 22.57 | A |
| #87 | 14.76 | 284.19 | 22.37 | A |
| #88 | 24.20 | 294.32 | 22.20 | A |
| #89 | 29.00 | 249.41 | 22.11 | A |
| #90 | 25.70 | 298.82 | 22.00 | A |
| #91 | 15.80 | 254.26 | 21.96 | A |
| #92 | 52.87 | 213.36 | 21.76 | A |
| #93 | 29.69 | 291.63 | 21.71 | A |
| #94 | −10.13 | 317.21 | 21.60 | C |
| #95 | 45.39 | 232.33 | 21.50 | A |
| #96 | 38.77 | 198.37 | 21.36 | A |
| #97 | −48.47 | 320.64 | 21.34 | A |
| #98 | −18.61 | 309.79 | 21.26 | B |
| #99 | −18.65 | 224.61 | 21.18 | A |
| #100 | −12.69 | 326.68 | 21.07 | A |
| #101 | −20.74 | 313.64 | 21.01 | A |
| #102 | −17.96 | 221.27 | 20.96 | A |
| #103 | 39.12 | 195.69 | 20.92 | A |
| #104 | −19.34 | 334.24 | 20.85 | A |
| #105 | 57.09 | 218.69 | 20.85 | C |
| #106 | 11.70 | 265.11 | 20.80 | A |
| #107 | 11.28 | 255.01 | 20.61 | A |
| #108 | 56.59 | 188.84 | 20.60 | A |
| #109 | 22.02 | 265.29 | 20.55 | A |
| #110 | 21.39 | 243.42 | 20.51 | A |
| #111 | −5.22 | 293.23 | 20.49 | A |
| #112 | 14.70 | 321.15 | 20.43 | A |
| #113 | 21.41 | 277.08 | 20.33 | A |
| #114 | 42.91 | 217.22 | 20.31 | A |
| #115 | 33.20 | 247.58 | 20.09 | A |
| #116 | 30.60 | 262.68 | 20.07 | A |
| #117 | 12.80 | 243.55 | 19.95 | B |
| #118 | 12.68 | 306.65 | 19.75 | B |
| #119 | 71.37 | 207.95 | 19.75 | C |
| #120 | 21.89 | 298.77 | 19.75 | A |
| #121 | 20.33 | 252.26 | 19.73 | B |
| #122 | 31.00 | 288.78 | 19.71 | A |
| #123 | 30.73 | 254.88 | 19.68 | C |
| #124 | 10.14 | 253.84 | 19.67 | A |
| #125 | −27.41 | 355.70 | 19.65 | A |
| #126 | 18.77 | 275.89 | 19.62 | A |
| #127 | −16.80 | 328.05 | 19.59 | B |
| #128 | −28.55 | 317.09 | 19.47 | A |
| #129 | 20.91 | 304.40 | 19.38 | A |
| #130 | 22.09 | 291.16 | 19.29 | A |
| #131 | 11.01 | 301.10 | 19.24 | A |
| #132 | 56.36 | 232.52 | 19.19 | A |
| #133 | −51.63 | 327.24 | 19.15 | A |
| #134 | 7.93 | 300.81 | 19.11 | A |
| #135 | 13.91 | 336.51 | 19.09 | A |
| #136 | −8.73 | 309.23 | 18.96 | A |
| #137 | 17.45 | 307.52 | 18.75 | A |
| #138 | −51.92 | 336.00 | 18.75 | B |
| #139 | 9.21 | 250.31 | 18.73 | A |
| #140 | −35.76 | 295.93 | 18.73 | A |
| #141 | 42.11 | 220.59 | 18.71 | A |
| #142 | 20.77 | 282.79 | 18.47 | A |
| #143 | −11.80 | 292.97 | 18.46 | A |
| #144 | 16.30 | 259.35 | 18.42 | A |
| #145 | 30.50 | 251.99 | 18.41 | A |
| #146 | −48.29 | 343.59 | 18.40 | A |
| #147 | 20.78 | 297.98 | 18.37 | A |
| #148 | 42.96 | 199.33 | 18.25 | A |
| #149 | 17.01 | 305.99 | 18.24 | A |
| #150 | 23.14 | 243.19 | 18.17 | A |
| #151 | 37.13 | 298.36 | 18.01 | A |
| #152 | 13.22 | 301.54 | 17.98 | A |
| #153 | 24.74 | 259.17 | 17.97 | C |
| #154 | 15.11 | 245.24 | 17.95 | C |
| #155 | −14.81 | 233.43 | 17.94 | C |
| #156 | 9.20 | 294.18 | 17.90 | A |
| #157 | −37.71 | 311.98 | 17.88 | A |
| #158 | 7.50 | 334.13 | 17.71 | A |
| #159 | 67.58 | 227.43 | 17.65 | C |
| #160 | 39.95 | 257.90 | 17.64 | A |
| #161 | −9.45 | 294.02 | 17.54 | C |
| #162 | −32.94 | 295.06 | 17.50 | A |
| #163 | 23.14 | 296.70 | 17.46 | A |
| #164 | 30.70 | 275.24 | 17.45 | A |
| #165 | 40.70 | 217.71 | 17.33 | A |
| #166 | 48.47 | 309.83 | 17.28 | A |
| #167 | 6.50 | 326.14 | 17.16 | C |
| #168 | 37.54 | 296.23 | 17.13 | A |
| #169 | −18.71 | 230.20 | 16.93 | C |
| #170 | −25.91 | 321.34 | 16.88 | A |
| #171 | −22.72 | 321.18 | 16.86 | A |
| #172 | 17.68 | 277.25 | 16.85 | A |
| #173 | 37.02 | 222.21 | 16.76 | A |
| #174 | 11.55 | 292.57 | 16.71 | A |
| #175 | 36.02 | 261.17 | 16.68 | A |
| #176 | 25.17 | 238.63 | 16.50 | A |
| #177 | −41.34 | 332.54 | 16.48 | A |
| #178 | −37.20 | 302.76 | 16.43 | A |
| #179 | −0.36 | 295.97 | 16.33 | A |
| #180 | 4.95 | 302.89 | 16.23 | A |
| #181 | −7.76 | 306.04 | 16.19 | A |
| #182 | 38.73 | 253.63 | 16.11 | A |
| #183 | 6.55 | 244.01 | 16.01 | B |
| #184 | −16.97 | 233.01 | 15.98 | A |
| #185 | 48.42 | 188.53 | 15.93 | B |
| #186 | 3.44 | 303.00 | 15.88 | A |
| #187 | −70.59 | 343.01 | 15.85 | A |
| #188 | 41.04 | 300.94 | 15.83 | C |
| #189 | −58.31 | 347.31 | 15.82 | A |
| #190 | 7.51 | 323.98 | 15.68 | A |
| #191 | 40.94 | 316.49 | 15.60 | A |
| #192 | −12.30 | 310.53 | 15.52 | A |
| #193 | 17.95 | 257.73 | 15.51 | B |
| #194 | −36.47 | 347.38 | 15.51 | A |
| #195 | 18.68 | 280.90 | 15.45 | A |
| #196 | −52.19 | 316.39 | 15.38 | A |
| #197 | −10.12 | 319.97 | 15.36 | C |
| #198 | −30.57 | 347.80 | 15.27 | A |
| #199 | 54.79 | 225.79 | 15.25 | A |
| #200 | 23.97 | 262.06 | 15.21 | C |
| #201 | 15.38 | 261.12 | 15.18 | A |
| #202 | 19.19 | 303.53 | 15.14 | A |
| #203 | 27.21 | 259.76 | 15.10 | C |
| #204 | −3.93 | 312.70 | 15.10 | A |
| #205 | 25.46 | 291.76 | 15.09 | A |
| #206 | −55.50 | 330.37 | 15.01 | A |
| #207 | 17.79 | 302.41 | 14.96 | A |
| #208 | 32.01 | 256.76 | 14.96 | A |
| #209 | 14.98 | 296.09 | 14.95 | A |
| #210 | −13.16 | 316.26 | 14.86 | A |
| #211 | 35.32 | 303.16 | 14.85 | A |
| #212 | −33.96 | 354.48 | 14.78 | A |
| #213 | 14.26 | 328.15 | 14.75 | A |
| #214 | −18.32 | 293.34 | 14.73 | C |
| #215 | 38.16 | 270.93 | 14.72 | A |
| #216 | −48.74 | 333.64 | 14.71 | C |
| #217 | 15.10 | 330.78 | 14.71 | A |
| #218 | 33.81 | 274.26 | 14.71 | A |
| #219 | 40.13 | 202.59 | 14.65 | C |
| #220 | −48.71 | 345.57 | 14.65 | B |
| #221 | 7.91 | 244.93 | 14.63 | A |
| #222 | 34.48 | 251.33 | 14.58 | A |
| #223 | −25.08 | 327.06 | 14.56 | A |
| #224 | −8.46 | 316.19 | 14.45 | C |
| #225 | −21.18 | 219.03 | 14.41 | A |
| #226 | 17.96 | 260.68 | 14.32 | C |
| #227 | 10.66 | 252.49 | 14.31 | A |
| #228 | −6.19 | 307.20 | 14.29 | A |
| #229 | 8.06 | 290.30 | 14.22 | A |
| #230 | −25.31 | 301.83 | 14.21 | A |
| #231 | −53.23 | 342.40 | 14.11 | A |
| #232 | 45.18 | 205.96 | 14.10 | A |
| #233 | −9.24 | 320.85 | 14.09 | A |
| #234 | 35.47 | 264.43 | 14.03 | A |
| #235 | −34.07 | 307.38 | 13.96 | A |
| #236 | 33.60 | 253.46 | 13.96 | A |
| #237 | 7.56 | 327.30 | 13.95 | A |
| #238 | −26.19 | 354.53 | 13.94 | A |
| #239 | −6.36 | 295.08 | 13.90 | B |
| #240 | 19.70 | 301.90 | 13.89 | A |
| #241 | 18.98 | 292.02 | 13.89 | A |
| #242 | −35.06 | 298.78 | 13.86 | A |
| #243 | −27.60 | 337.55 | 13.83 | A |
| #244 | −12.07 | 308.60 | 13.78 | A |
| #245 | −26.18 | 312.80 | 13.77 | B |
| #246 | −47.93 | 336.21 | 13.74 | C |
| #247 | 33.44 | 298.93 | 13.67 | A |
| #248 | 18.45 | 300.30 | 13.64 | A |
| #249 | 35.02 | 245.58 | 13.63 | A |
| #250 | 13.85 | 325.09 | 13.59 | A |
| #251 | 33.54 | 263.49 | 13.53 | A |
| #252 | 26.91 | 247.50 | 13.51 | B |
| #253 | 6.74 | 257.82 | 13.44 | A |
| #254 | 51.06 | 182.56 | 13.41 | A |
| #255 | 40.98 | 201.75 | 13.39 | A |
| #256 | −2.02 | 310.82 | 13.37 | A |
| #257 | −21.27 | 327.10 | 13.36 | A |
| #258 | 45.56 | 236.29 | 13.26 | A |
| #259 | 8.06 | 250.91 | 13.26 | A |
| #260 | −30.97 | 339.76 | 13.07 | A |
| #261 | −29.12 | 336.42 | 13.07 | A |
| #262 | 15.30 | 257.51 | 13.06 | A |
| #263 | −23.58 | 312.96 | 13.06 | A |
| #264 | 21.65 | 245.27 | 13.05 | A |
| #265 | 21.12 | 279.02 | 13.00 | A |
| #266 | 16.41 | 287.74 | 12.99 | C |
| #267 | 14.03 | 261.74 | 12.95 | A |
| #268 | 12.30 | 324.69 | 12.95 | A |
| #269 | 9.76 | 265.74 | 12.90 | A |
| #270 | −35.94 | 310.30 | 12.90 | A |
| #271 | −23.11 | 319.42 | 12.86 | A |
| #272 | −50.61 | 336.86 | 12.84 | A |
| #273 | −20.24 | 216.39 | 12.83 | C |
| #274 | 14.75 | 319.78 | 12.76 | A |
| #275 | −21.88 | 312.39 | 12.75 | A |
| #276 | −6.84 | 315.16 | 12.72 | B |
| #277 | 33.97 | 265.53 | 12.69 | A |
| #278 | −18.60 | 328.19 | 12.68 | A |
| #279 | 25.50 | 254.62 | 12.64 | A |
| #280 | 40.76 | 265.68 | 12.64 | A |
| #281 | 15.90 | 246.50 | 12.58 | A |
| #282 | −22.49 | 329.80 | 12.50 | A |
| #283 | 17.86 | 242.87 | 12.50 | A |
| #284 | 41.07 | 251.01 | 12.47 | A |
| #285 | −36.13 | 351.06 | 12.47 | A |
| #286 | 13.84 | 256.02 | 12.46 | A |
| #287 | 41.69 | 194.72 | 12.42 | B |
| #288 | −68.48 | 344.44 | 12.37 | A |
| #289 | 2.74 | 244.41 | 12.32 | A |
| #290 | 7.43 | 329.76 | 12.27 | A |
| #291 | 16.57 | 284.19 | 12.27 | C |
| #292 | 35.47 | 296.88 | 12.27 | B |
| #293 | −11.45 | 312.35 | 12.22 | A |
| #294 | 50.56 | 166.62 | 12.20 | A |
| #295 | 30.77 | 248.82 | 12.20 | A |
| #296 | 44.33 | 202.23 | 12.14 | A |
| #297 | 68.04 | 204.06 | 12.14 | A |
| #298 | −32.09 | 347.94 | 12.09 | A |
| #299 | 21.24 | 255.59 | 12.07 | A |
| #300 | −38.25 | 353.98 | 12.07 | A |
| #301 | 40.05 | 255.65 | 12.06 | A |
| #302 | 19.99 | 239.90 | 12.04 | A |
| #303 | 24.44 | 248.23 | 11.99 | A |
| #304 | −24.83 | 352.37 | 11.99 | A |
| #305 | 41.37 | 253.47 | 11.98 | A |
| #306 | 8.62 | 261.29 | 11.96 | A |
| #307 | 8.55 | 262.37 | 11.94 | A |
| #308 | 35.37 | 243.52 | 11.92 | A |
| #309 | 15.15 | 293.55 | 11.91 | A |
| #310 | 45.29 | 234.71 | 11.90 | A |
| #311 | 55.11 | 161.99 | 11.84 | C |
| #312 | 25.97 | 251.20 | 11.81 | A |
| #313 | 25.70 | 244.63 | 11.75 | A |
| #314 | 17.50 | 285.53 | 11.71 | A |
| #315 | 37.22 | 264.88 | 11.70 | A |
| #316 | −2.09 | 286.24 | 11.69 | B |
| #317 | −26.93 | 325.69 | 11.67 | A |
| #318 | 11.80 | 253.53 | 11.60 | A |
| #319 | 10.36 | 332.82 | 11.52 | A |
| #320 | 39.00 | 256.26 | 11.51 | A |
| #321 | 42.79 | 226.39 | 11.50 | A |
| #322 | 47.33 | 152.82 | 11.49 | C |
| #323 | 49.34 | 157.79 | 11.49 | A |
| #324 | 17.48 | 294.47 | 11.44 | A |
| #325 | 38.79 | 204.38 | 11.42 | A |
| #326 | 10.18 | 249.71 | 11.40 | A |
| #327 | 21.85 | 239.99 | 11.38 | A |
| #328 | −18.38 | 322.47 | 11.36 | A |
| #329 | 41.37 | 315.37 | 11.34 | C |
| #330 | 4.20 | 254.19 | 11.30 | A |
| #331 | 41.27 | 204.00 | 11.28 | A |
| #332 | 25.12 | 288.53 | 11.25 | B |
| #333 | 54.98 | 187.12 | 11.23 | C |
| #334 | −55.09 | 332.31 | 11.20 | A |
| #335 | −24.62 | 298.16 | 11.19 | C |
| #336 | −53.37 | 347.23 | 11.12 | C |
| #337 | 9.81 | 313.30 | 11.11 | A |
| #338 | 41.95 | 253.96 | 11.10 | C |
| #339 | 32.18 | 297.04 | 11.05 | A |
| #340 | 40.34 | 204.32 | 11.04 | A |
| #341 | −19.92 | 322.69 | 10.98 | A |
| #342 | −35.56 | 306.20 | 10.98 | A |
| #343 | 69.64 | 200.36 | 10.97 | C |
| #344 | 39.38 | 213.39 | 10.94 | A |
| #345 | 35.54 | 297.88 | 10.94 | C |
| #346 | 13.25 | 302.99 | 10.92 | A |
| #347 | 47.30 | 219.07 | 10.91 | A |
| #348 | 11.16 | 294.51 | 10.91 | A |
| #349 | 14.97 | 248.90 | 10.86 | C |
| #350 | 12.58 | 251.03 | 10.86 | A |
| #351 | 5.74 | 254.40 | 10.86 | B |
| #352 | 22.14 | 255.61 | 10.82 | C |
| #353 | −58.56 | 353.19 | 10.82 | A |
| #354 | 54.52 | 186.14 | 10.76 | A |
| #355 | 7.37 | 331.29 | 10.75 | A |
| #356 | 8.75 | 248.62 | 10.75 | A |
| #357 | 52.43 | 183.97 | 10.70 | A |
| #358 | −17.94 | 327.76 | 10.70 | C |
| #359 | 23.82 | 239.36 | 10.69 | A |
| #360 | 6.83 | 272.06 | 10.68 | A |
| #361 | 16.92 | 283.56 | 10.65 | C |
| #362 | 34.19 | 258.98 | 10.61 | A |
| #363 | −48.96 | 312.37 | 10.61 | A |
| #364 | 43.19 | 311.70 | 10.57 | A |
| #365 | 12.75 | 258.30 | 10.54 | A |
| #366 | 43.27 | 207.21 | 10.53 | A |
| #367 | 13.45 | 296.32 | 10.52 | A |
| #368 | −19.50 | 319.60 | 10.47 | A |
| #369 | −23.34 | 332.54 | 10.46 | A |
| #370 | −23.86 | 323.82 | 10.46 | A |
| #371 | 10.91 | 331.60 | 10.40 | A |
| #372 | 17.90 | 252.97 | 10.39 | A |
| #373 | 7.57 | 271.30 | 10.38 | A |
| #374 | 10.26 | 260.80 | 10.34 | A |
| #375 | 10.57 | 258.92 | 10.34 | A |
| #376 | 49.21 | 227.76 | 10.33 | B |
| #377 | 36.92 | 268.29 | 10.30 | A |
| #378 | −18.64 | 330.45 | 10.26 | A |
| #379 | 23.95 | 302.76 | 10.25 | A |
| #380 | 6.85 | 328.06 | 10.20 | A |
| #381 | 9.75 | 270.39 | 10.10 | C |
| #382 | 9.89 | 264.13 | 10.07 | A |
| #383 | 8.77 | 323.34 | 10.06 | A |
| #384 | 4.65 | 297.36 | 10.05 | A |
| #385 | 45.36 | 214.89 | 10.03 | A |
| #386 | 10.81 | 264.83 | 10.02 | A |
| #387 | 65.46 | 214.49 | 10.00 | A |
| #388 | 27.53 | 261.41 | 9.97 | A |
| #389 | 38.05 | 259.78 | 9.94 | A |
| #390 | 15.63 | 323.09 | 9.93 | B |
| #391 | 26.64 | 249.95 | 9.91 | A |
| #392 | 11.95 | 260.79 | 9.90 | B |
| #393 | −18.63 | 312.10 | 9.89 | A |
| #394 | 13.87 | 251.65 | 9.87 | A |
| #395 | 65.81 | 228.53 | 9.80 | B |
| #396 | 36.95 | 306.37 | 9.75 | A |
| #397 | 2.50 | 254.31 | 9.75 | A |
| #398 | −19.36 | 321.59 | 9.71 | A |
| #399 | 12.09 | 246.72 | 9.69 | A |
| #400 | 40.85 | 222.48 | 9.66 | A |
| #401 | 24.97 | 302.28 | 9.66 | A |
| #402 | −60.52 | 334.17 | 9.64 | B |
| #403 | 5.66 | 257.49 | 9.63 | C |
| #404 | −43.26 | 334.54 | 9.60 | A |
| #405 | 33.33 | 250.13 | 9.59 | C |
| #406 | 25.32 | 264.55 | 9.57 | A |
| #407 | 45.29 | 203.23 | 9.53 | A |
| #408 | −35.04 | 301.09 | 9.52 | A |
| #409 | 28.48 | 252.90 | 9.50 | A |
| #410 | 9.52 | 318.16 | 9.49 | A |
| #411 | 8.92 | 259.58 | 9.46 | A |
| #412 | 11.51 | 273.33 | 9.43 | A |
| #413 | 28.51 | 265.42 | 9.41 | A |
| #414 | 65.90 | 223.98 | 9.38 | A |
| #415 | 32.92 | 293.05 | 9.37 | A |
| #416 | −47.12 | 312.43 | 9.36 | A |
| #417 | 40.16 | 215.35 | 9.33 | A |
| #418 | −44.16 | 306.21 | 9.31 | A |
| #419 | −26.47 | 317.86 | 9.28 | A |
| #420 | −48.35 | 311.59 | 9.25 | C |
| #421 | −53.25 | 339.73 | 9.20 | A |
| #422 | 10.19 | 275.32 | 9.19 | C |
| #423 | 11.77 | 298.43 | 9.18 | A |
| #424 | 18.25 | 241.51 | 9.17 | A |
| #425 | 37.77 | 259.06 | 9.17 | A |
| #426 | 13.53 | 260.25 | 9.14 | C |
| #427 | 16.52 | 244.87 | 9.14 | A |
| #428 | 24.99 | 297.23 | 9.13 | A |
| #429 | −24.83 | 313.91 | 9.12 | A |
| #430 | 14.21 | 268.02 | 9.12 | A |
| #431 | 33.73 | 251.75 | 9.10 | A |
| #432 | 22.04 | 292.80 | 9.05 | A |
| #433 | 36.87 | 273.13 | 9.04 | A |
| #434 | −49.32 | 321.03 | 9.02 | A |
| #435 | 19.41 | 283.76 | 9.00 | A |
| #436 | −15.46 | 309.41 | 8.99 | A |
| #437 | 44.75 | 311.49 | 8.98 | A |
| #438 | 22.98 | 255.11 | 8.95 | C |
| #439 | −25.19 | 317.26 | 8.94 | A |
| #440 | 21.58 | 241.40 | 8.91 | A |
| #441 | 17.73 | 245.91 | 8.87 | A |
| #442 | 12.44 | 259.48 | 8.87 | A |
| #443 | 15.15 | 258.57 | 8.83 | C |
| #444 | 6.23 | 260.44 | 8.82 | C |
| #445 | 36.35 | 255.26 | 8.79 | A |
| #446 | 18.92 | 305.81 | 8.78 | A |
| #447 | 37.43 | 258.02 | 8.73 | A |
| #448 | 19.46 | 246.45 | 8.72 | A |
| #449 | 10.56 | 324.60 | 8.69 | A |
| #450 | 8.49 | 296.34 | 8.67 | B |
| #451 | 39.79 | 304.48 | 8.63 | B |
| #452 | 14.99 | 267.62 | 8.59 | A |
| #453 | 8.63 | 260.29 | 8.59 | A |
| #454 | 13.14 | 252.02 | 8.50 | A |
| #455 | −53.01 | 321.61 | 8.48 | A |
| #456 | 29.27 | 261.48 | 8.48 | A |
| #457 | 12.65 | 311.80 | 8.46 | C |
| #458 | 41.19 | 200.22 | 8.45 | C |
| #459 | 11.23 | 279.21 | 8.44 | A |
| #460 | 11.96 | 252.32 | 8.43 | A |
| #461 | 8.43 | 268.15 | 8.43 | A |
| #462 | 10.86 | 256.49 | 8.42 | A |
| #463 | −39.10 | 304.31 | 8.41 | B |
| #464 | −58.97 | 324.65 | 8.41 | A |
| #465 | 42.37 | 255.48 | 8.35 | A |
| #466 | 28.09 | 260.26 | 8.35 | A |
| #467 | 6.01 | 259.51 | 8.35 | A |
| #468 | 8.92 | 268.11 | 8.28 | A |
| #469 | 8.11 | 322.69 | 8.27 | A |
| #470 | 19.69 | 254.73 | 8.18 | A |
| #471 | 33.68 | 291.89 | 8.17 | A |
| #472 | −46.52 | 311.52 | 8.13 | A |
| #473 | 41.39 | 199.26 | 8.13 | C |
| #474 | 31.90 | 260.40 | 8.04 | C |
| #475 | −46.91 | 334.56 | 8.02 | A |
| #476 | 16.22 | 304.45 | 8.01 | |
| #477 | −59.14 | 331.10 | 8.00 | |
| #478 | 44.04 | 311.16 | 7.95 | |
| #479 | 36.78 | 287.10 | 7.90 | |
| #480 | 17.84 | 255.23 | 7.83 | |
| #481 | 44.31 | 217.49 | 7.81 | |
| #482 | 6.04 | 252.32 | 7.80 | |
| #483 | 62.91 | 235.59 | 7.78 | |
| #484 | 41.54 | 214.62 | 7.76 | |
| #485 | 49.52 | 163.78 | 7.74 | |

| | | | |
|---|---|---|---|
| #486 | 31.42 | 244.98 | 7.74 |
| #487 | 10.79 | 329.03 | 7.74 |
| #488 | −33.13 | 345.30 | 7.71 |
| #489 | 62.35 | 223.24 | 7.69 |
| #490 | 8.59 | 257.28 | 7.67 |
| #491 | 49.45 | 314.41 | 7.67 |
| #492 | 41.23 | 261.11 | 7.67 |
| #493 | 35.54 | 305.34 | 7.64 |
| #494 | −56.05 | 336.97 | 7.64 |
| #495 | 39.61 | 266.23 | 7.63 |
| #496 | 49.07 | 209.90 | 7.63 |
| #497 | 13.33 | 268.43 | 7.63 |
| #498 | 11.90 | 301.55 | 7.61 |
| #499 | 13.24 | 284.81 | 7.61 |
| #500 | 21.54 | 294.57 | 7.60 |
| #501 | −29.83 | 208.90 | 7.59 |
| #502 | 11.47 | 336.13 | 7.59 |
| #503 | 10.53 | 267.89 | 7.58 |
| #504 | −32.85 | 302.89 | 7.58 |
| #505 | −56.38 | 339.07 | 7.54 |
| #506 | 10.65 | 268.81 | 7.51 |
| #507 | 8.81 | 319.63 | 7.49 |
| #508 | 10.21 | 257.91 | 7.49 |
| #509 | 40.07 | 267.56 | 7.48 |
| #510 | −36.77 | 308.14 | 7.48 |
| #511 | 50.27 | 202.35 | 7.45 |
| #512 | −21.12 | 221.85 | 7.45 |
| #513 | 48.19 | 158.10 | 7.45 |
| #514 | 25.92 | 251.18 | 7.41 |
| #515 | 52.88 | 187.20 | 7.40 |
| #516 | 41.35 | 223.61 | 7.38 |
| #517 | 9.53 | 268.93 | 7.37 |
| #518 | 6.94 | 332.25 | 7.36 |
| #519 | 35.74 | 306.42 | 7.36 |
| #520 | 11.05 | 267.43 | 7.34 |
| #521 | 14.88 | 298.57 | 7.32 |
| #522 | −35.55 | 345.24 | 7.28 |
| #523 | 12.04 | 251.59 | 7.27 |
| #524 | 42.15 | 197.45 | 7.25 |
| #525 | −29.02 | 210.52 | 7.24 |
| #526 | 38.40 | 266.37 | 7.24 |
| #527 | 32.14 | 253.48 | 7.21 |
| #528 | 39.15 | 271.40 | 7.21 |
| #529 | 14.54 | 260.23 | 7.19 |
| #530 | 46.36 | 314.31 | 7.19 |
| #531 | −29.07 | 209.31 | 7.18 |
| #532 | 12.46 | 262.53 | 7.18 |
| #533 | 47.12 | 313.34 | 7.18 |
| #534 | 17.18 | 255.96 | 7.17 |
| #535 | 13.09 | 254.59 | 7.15 |
| #536 | 53.97 | 183.32 | 7.14 |
| #537 | −41.29 | 309.42 | 7.10 |
| #538 | 14.72 | 303.15 | 7.09 |
| #539 | 43.89 | 222.80 | 7.07 |
| #540 | 49.52 | 235.35 | 7.07 |
| #541 | 16.03 | 261.80 | 6.99 |
| #542 | 31.70 | 249.28 | 6.99 |
| #543 | −48.80 | 322.67 | 6.97 |
| #544 | 9.08 | 332.37 | 6.96 |
| #545 | 5.45 | 258.74 | 6.96 |
| #546 | 9.36 | 325.99 | 6.94 |
| #547 | −59.33 | 338.57 | 6.93 |
| #548 | 15.56 | 253.79 | 6.91 |
| #549 | 8.37 | 321.61 | 6.90 |
| #550 | 24.44 | 250.17 | 6.89 |
| #551 | 25.81 | 248.35 | 6.88 |
| #552 | −7.10 | 307.71 | 6.87 |
| #553 | 34.68 | 291.59 | 6.85 |
| #554 | 40.33 | 209.01 | 6.84 |
| #555 | 15.68 | 319.08 | 6.76 |
| #556 | 49.76 | 202.30 | 6.72 |
| #557 | 46.02 | 216.84 | 6.70 |
| #558 | 26.82 | 296.91 | 6.70 |
| #559 | 1.79 | 251.33 | 6.70 |
| #560 | 13.47 | 257.97 | 6.70 |
| #561 | −57.42 | 329.90 | 6.68 |
| #562 | −49.19 | 319.46 | 6.67 |
| #563 | 41.06 | 264.49 | 6.64 |
| #564 | 1.31 | 256.11 | 6.64 |
| #565 | 28.51 | 254.42 | 6.64 |
| #566 | 46.02 | 212.81 | 6.62 |
| #567 | −40.04 | 301.03 | 6.59 |
| #568 | 38.24 | 269.77 | 6.59 |
| #569 | 36.89 | 301.36 | 6.58 |
| #570 | 28.71 | 295.20 | 6.56 |
| #571 | 29.43 | 296.07 | 6.55 |
| #572 | 8.11 | 253.78 | 6.54 |
| #573 | 10.80 | 270.83 | 6.53 |
| #574 | 17.28 | 259.22 | 6.50 |
| #575 | 47.78 | 217.31 | 6.50 |
| #576 | 50.04 | 230.74 | 6.50 |
| #577 | 10.37 | 266.75 | 6.50 |
| #578 | 14.58 | 263.47 | 6.50 |
| #579 | −22.88 | 220.84 | 6.45 |
| #580 | 16.48 | 302.94 | 6.43 |
| #581 | 10.20 | 263.11 | 6.40 |
| #582 | 36.01 | 272.94 | 6.40 |
| #583 | 30.70 | 247.72 | 6.36 |
| #584 | 10.14 | 267.48 | 6.36 |
| #585 | 26.93 | 251.13 | 6.32 |
| #586 | 32.06 | 265.54 | 6.29 |
| #587 | 5.50 | 250.24 | 6.29 |
| #588 | 19.44 | 300.22 | 6.29 |
| #589 | 55.69 | 165.28 | 6.28 |
| #590 | 34.82 | 262.70 | 6.28 |
| #591 | −64.45 | 329.24 | 6.27 |
| #592 | 37.55 | 253.46 | 6.25 |
| #593 | 10.20 | 322.35 | 6.25 |
| #594 | −46.45 | 339.04 | 6.21 |
| #595 | −46.11 | 337.87 | 6.19 |
| #596 | 24.88 | 256.80 | 6.18 |
| #597 | −57.24 | 341.87 | 6.18 |
| #598 | 36.60 | 253.28 | 6.17 |
| #599 | 13.58 | 262.73 | 6.16 |
| #600 | 26.27 | 265.30 | 6.15 |
| #601 | 38.75 | 237.18 | 6.12 |
| #602 | 42.13 | 202.94 | 6.08 |
| #603 | 26.29 | 254.61 | 6.06 |
| #604 | 28.85 | 256.29 | 6.06 |
| #605 | 13.84 | 304.54 | 6.04 |
| #606 | 38.82 | 307.15 | 6.04 |
| #607 | 7.95 | 255.83 | 6.00 |
| #608 | 38.96 | 305.24 | 5.97 |
| #609 | 8.65 | 328.37 | 5.92 |
| #610 | 28.49 | 298.46 | 5.92 |
| #611 | 33.92 | 255.65 | 5.87 |
| #612 | 70.21 | 212.88 | 5.87 |
| #613 | 67.50 | 206.94 | 5.86 |
| #614 | −58.37 | 328.89 | 5.86 |
| #615 | 16.06 | 296.83 | 5.86 |
| #616 | 43.45 | 227.68 | 5.85 |
| #617 | −17.73 | 326.13 | 5.84 |
| #618 | 14.14 | 259.21 | 5.84 |
| #619 | 36.76 | 305.33 | 5.84 |
| #620 | 56.17 | 224.90 | 5.83 |
| #621 | 28.11 | 258.21 | 5.82 |
| #622 | 44.81 | 315.88 | 5.81 |
| #623 | 35.66 | 304.40 | 5.80 |
| #624 | −44.01 | 326.70 | 5.80 |
| #625 | 1.92 | 255.48 | 5.78 |
| #626 | 41.04 | 211.97 | 5.74 |
| #627 | 11.19 | 245.15 | 5.72 |
| #628 | 7.37 | 256.35 | 5.67 |
| #629 | 30.39 | 298.05 | 5.66 |
| #630 | 14.42 | 302.20 | 5.61 |
| #631 | 15.08 | 253.60 | 5.61 |
| #632 | 39.05 | 260.86 | 5.58 |
| #633 | 42.22 | 305.44 | 5.56 |
| #634 | 50.16 | 234.72 | 5.54 |
| #635 | 30.47 | 271.44 | 5.54 |
| #636 | 62.06 | 234.86 | 5.52 |
| #637 | 7.22 | 251.80 | 5.51 |
| #638 | 40.53 | 209.40 | 5.48 |
| #639 | 52.18 | 166.74 | 5.47 |
| #640 | 52.51 | 167.46 | 5.46 |
| #641 | 35.11 | 259.45 | 5.45 |
| #642 | 6.96 | 332.97 | 5.44 |
| #643 | 16.18 | 257.64 | 5.44 |
| #644 | 52.12 | 165.46 | 5.44 |
| #645 | 11.66 | 256.56 | 5.41 |
| #646 | 20.79 | 300.62 | 5.40 |
| #647 | −57.79 | 343.90 | 5.37 |
| #648 | −15.82 | 233.69 | 5.37 |
| #649 | −8.14 | 311.63 | 5.35 |
| #650 | 43.76 | 256.65 | 5.34 |
| #651 | 30.80 | 238.24 | 5.34 |
| #652 | 22.36 | 238.64 | 5.33 |
| #653 | 2.92 | 249.35 | 5.33 |
| #654 | 8.44 | 329.12 | 5.33 |
| #655 | −11.35 | 311.91 | 5.32 |
| #656 | 20.54 | 264.61 | 5.29 |
| #657 | 54.42 | 227.07 | 5.29 |
| #658 | 36.69 | 269.69 | 5.28 |
| #659 | 8.83 | 329.89 | 5.25 |
| #660 | 11.73 | 302.51 | 5.24 |
| #661 | 35.95 | 266.26 | 5.22 |
| #662 | 47.62 | 222.23 | 5.17 |
| #663 | 46.25 | 233.45 | 5.16 |
| #664 | 12.93 | 245.24 | 5.13 |
| #665 | 29.58 | 240.25 | 5.13 |
| #666 | −0.33 | 247.72 | 5.10 |
| #667 | 38.58 | 263.37 | 5.08 |
| #668 | 25.54 | 298.99 | 5.07 |
| #669 | 67.55 | 221.06 | 5.05 |
| #670 | 42.19 | 250.69 | 5.01 |
| #671 | 18.76 | 242.45 | 4.97 |
| #672 | 41.99 | 237.49 | 4.94 |
| #673 | 12.66 | 261.57 | 4.93 |
| #674 | 42.86 | 254.21 | 4.93 |
| #675 | 9.37 | 321.69 | 4.86 |
| #676 | 2.53 | 250.61 | 4.85 |
| #677 | 2.22 | 246.32 | 4.82 |
| #678 | 13.61 | 254.81 | 4.80 |
| #679 | 14.43 | 298.89 | 4.79 |
| #680 | 9.83 | 261.64 | 4.78 |
| #681 | −38.04 | 310.05 | 4.76 |
| #682 | 58.09 | 235.07 | 4.74 |
| #683 | 38.16 | 223.45 | 4.72 |
| #684 | 18.02 | 246.82 | 4.72 |
| #685 | −57.04 | 340.21 | 4.72 |
| #686 | 14.19 | 302.68 | 4.70 |
| #687 | 35.88 | 257.74 | 4.66 |
| #688 | 9.86 | 324.32 | 4.65 |
| #689 | 13.77 | 252.24 | 4.64 |
| #690 | 15.41 | 300.98 | 4.55 |
| #691 | 0.22 | 247.64 | 4.53 |
| #692 | 6.93 | 259.85 | 4.49 |
| #693 | 44.66 | 258.66 | 4.47 |
| #694 | 31.83 | 250.63 | 4.42 |
| #695 | −44.48 | 327.01 | 4.40 |
| #696 | 0.62 | 246.91 | 4.38 |
| #697 | −17.07 | 291.44 | 4.29 |
| #698 | 0.02 | 246.84 | 4.27 |
| #699 | 49.81 | 163.48 | 4.24 |
| #700 | 40.08 | 212.55 | 4.20 |